\documentclass[journal,a4paper]{IEEEtran}

\pdfoutput=1

\usepackage{graphicx}
\usepackage{amssymb, amsmath, amsfonts, amsthm}
\usepackage{multirow}
\usepackage{bm}
\usepackage{makecell}
\usepackage{nomencl}
\usepackage{mcite}
\usepackage[noadjust]{cite}
\usepackage{setspace}
\usepackage{flushend}
\usepackage{xcolor}

\allowdisplaybreaks

\IEEEoverridecommandlockouts

\begin{document}

\title{Recursive Secondary Controller for \\ Voltage Profile Improvement Based on \\ Primary Virtual Admittance Control
\thanks{D.~Moutevelis, J. Rold\'{a}n-P\'{e}rez, N.~Jankovic, and M. Prodanovic are with the Electrical Systems Unit, IMDEA Energy, Avda. Ram\'{o}n de la Sagra 3, 28935, M\'{o}stoles, Madrid, Spain.  E-mail:
    dionysios.moutevelis@imdea.org, javier.roldan@imdea.org, njegos.jankovic@imdea.org, milan.prodanovic@imdea.org.
    \\
    D.~Moutevelis is also with Alcal\'{a} de Henares University, Department of Electronics, Alcal\'{a} de Henares 28801, Madrid, Spain.
    }
\thanks{
This work has been partially financed by the Community of Madrid, through the research project PROMINT-CM, Ref: P2018/EMT4366, and by the Spanish Ministry of Science, Innovation and Universities through the Juan de la Cierva Incorporaci\'{o}n program (IJC2019-042342-I).
A shorter version of this paper was previously presented in a conference [1].
}}
\author{\IEEEauthorblockN{Dionysios Moutevelis, Javier Roldan-Perez, {\em Member, IEEE}, Njegos Jankovic, and \\ Milan Prodanovic, {\em Member, IEEE}}
}

\maketitle
\IEEEpubidadjcol
\begin{abstract}
This paper proposes a recursive, virtual admittance based, secondary controller for DG units that improves the voltage profile in distribution networks.
First, the adaptation of the virtual admittance concept for the goal of voltage regulation is explained.
Then, a recursive secondary controller is developed to periodically update the virtual admittance gains.
The controller is formulated as an optimization problem with current and stability limitations as constraints.
Measurements across the grid, transmitted through low-bandwidth communications, are used to simplify the calculations, resulting in a recursive algorithm.
Weight vectors are included in the objective function to allow participation flexibility of each converter.
Results show that the primary virtual admittance controller is able to mitigate over- and undervoltages in steady state and under transient conditions.
Subsequently, the secondary controller is shown to further improve the voltage profiles across the grid.
Experimental results obtained from a laboratory environment, comprising three DG units and a grid emulator, validate the functionality of the complete control structure.
\end{abstract}
\begin{IEEEkeywords} 
Virtual Admittance, Voltage Support, Decentralised Voltage Control, Distributed Generation. \phantom{\cite{moutevelis2021virtual}}
\end{IEEEkeywords}
\section{Introduction}
\label{sec.introduction}
In recent years, distributed generation (DG) has emerged as an efficient way for providing electricity in remote areas~\cite{moutevelis2021virtual,guerrero2010distributed}. 
DG allows the connection of renewable energy sources to low-voltage electrical networks and this is critical for promoting self consumption and local power generation.
At the same time, DG pose important technological challenges such as operating under transient over- and undervoltage conditions~\cite{carvalho2008distributed}.
In the past, DG units were expected to either preserve their normal operation during grid voltage variations or to disconnect in the case of severe disturbances.
However, the expectations and requests for them to actively contribute to the grid regulation has risen in recent years.
Specifically, voltage support is considered one of the main requirements for DG integration in electrical distribution systems~\cite{braun2012distribution}.

Several voltage regulating strategies have already been proposed for DG units.
The main distinction between them is whether they control active power, reactive power or a combination of the two.
They typically use a reactive power droop controller that takes advantage of the strong coupling between reactive power and voltage~\cite{vasquez2009voltage}.
However, in low-voltage networks the high $R/X$ ratio makes reactive power control a less effective solution for voltage regulation and active power management is also considered~\cite{tonkoski2010coordinated}.
To maximize voltage regulation capability of DG units, a combination of active and reactive power is usually performed both for undervoltage and overvoltage scenarios~\cite{collins2015real,fusco2021decentralized}.
The performance, effectiveness and impact of these strategies on the grid stability have already been addressed in the literature~\cite{braslavsky2017voltage}.
In all the studies mentioned above, active and reactive powers have been controlled independently. 
This means that the coupling between the two in low-voltage networks has been scantly explored and it is generally considered to be a side effect~\cite{zhong2022improving}.
Recently in~\cite{zhong2022improving}, $P-Q$ coupling was exploited in the control of distributed energy resources to improve the dynamic response of power systems.
However, transient conditions were prioritised over steady state voltage deviations and only simplified models for the current controllers and the power converters were used.

A different taxonomy of voltage control methods is based on their locality and architecture. The main categories are local, centralised and decentralised controllers as well as hybrid combinations of the above~\cite{evangelopoulos2016optimal}.
Local controllers use measurements only from the point of connection of the distributed generator and does not require any communication infrastructure.
However, historical data that might be unavailable are often used to tune the controllers~\cite{weckx2016optimal}.
Furthermore, it has been reported that corrective actions to improve the voltage profile locally can have an adverse effect in other nodes of the grid~\cite{fusco2021decentralized}.
Hence, a complementary control layer that monitors the voltage level of the whole network would be beneficial.
\begin{figure*}[!t]
\centering
\includegraphics[width=0.8\textwidth]{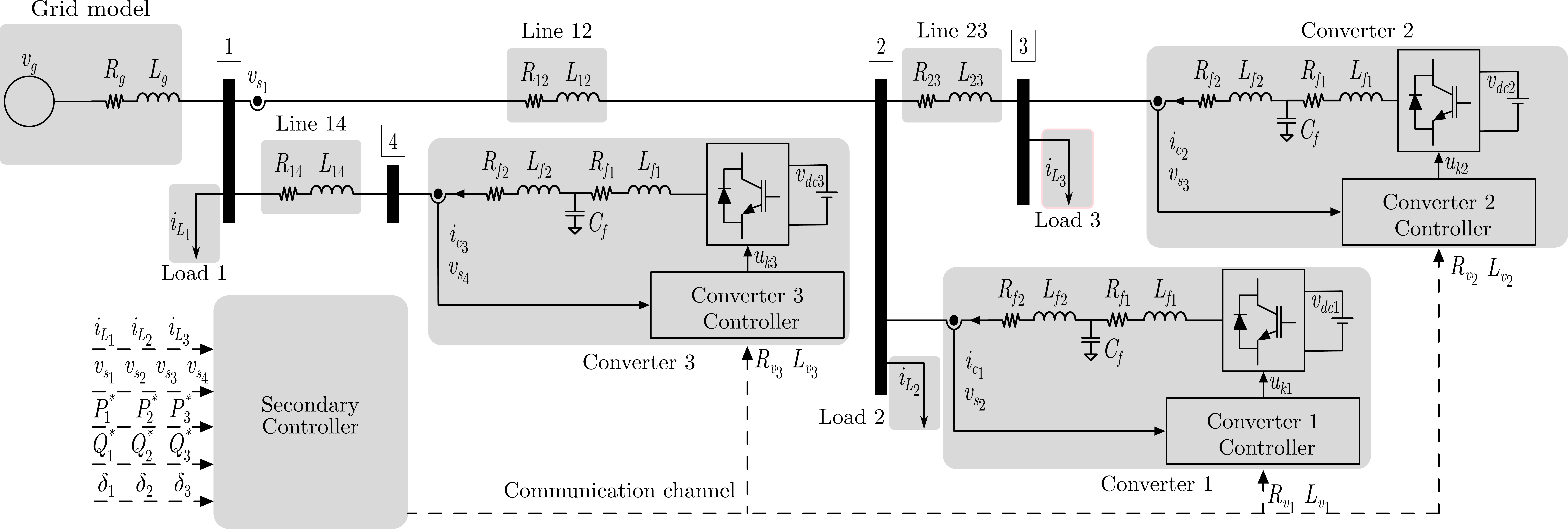}
\caption{Schematic of the electrical system studied in this work and overview of the two-level control configuration.}
\label{fig.grid}
\end{figure*} 

Centralised controllers use measurements from the whole grid and gives control set-points to each distributed generator~\cite{kim2012coordinated}. 
The main objective is to adjust the power injection to the system nodes in order to match the load consumption and thus, regulate their voltage.
The motivation is similar to the peak reduction and valley filling algorithms developed for electric vehicles~\cite{liang2018dynamic}.
Centralised methods allow the optimization of the grid operation.
However they require extensive grid metering and a robust communication network.
Also, they are vulnerable to cyber-attacks~\cite{fusco2021decentralized}. 
For this reason, a robust control layer that exclusively depends on local measurements is favourable. 

Decentralised controllers can refer to a wide variety of controllers that require minimal communication~\cite{gerdroodbari2021decentralized}.
Communication can be limited only to devices connected to the same phase, to devices connected to neighbouring nodes, or solely to low-bandwidth power line communication~\cite{olivier2015active}.
Their objectives may range from power loss reduction, overvoltage prevention or fairness in active power curtailment.
Lastly, hybrid approaches combining local and central control methods have been proposed~\cite{weckx2014combined}. 
Decentralised and hybrid methods take advantage of the robustness and reliability of local methods as well as make use of possible communication infrastructure to improve the performance of the whole network.
For this reason, they are considered a viable solution until fully monitored distribution grids with highly reliable, high bandwidth communication channels are available.

Current control is the fundamental piece of most grid-connected DG units. 
However, its negative impact on the grid stability has fostered the incorporation of circuit-inspired control loops in the control systems of DG units.
Specifically, the virtual impedance has been widely used to increase robustness and power-sharing among parallel DG units, mainly in microgrids, where DGs are operated as voltage sources~\cite{he2011analysis}. 
The dual concept of the virtual impedance for current-controlled converters is the virtual admittance. 
It was first introduced in~\cite{rodriguez2013control}, and it has been mainly used to emulate the equivalent inductance of synchronous machines~\cite{rodriguez2013control,rodriguez2018flexible,zhang2016frequency}
and for harmonic compensation~\cite{blanco2016virtual,micallef2015mitigation,gothner2021harmonic}.
This technique does not require derivatives of any signal, thus making its implementation straightforward and robust against noise~\cite{rodriguez2013control}. 
The motivation for applying the virtual admittance control for voltage profile improvement stems from the DG ability to operate seamlessly between inductive and capacitive modes while also adjusting its active power output.
This resembles an operation of a variable admittance and extends the capabilities of capacitor banks, a standard control measure against undervoltage in traditional power systems~\cite{aman2014optimum}.
This operation can contribute to balancing generation/load consumption and thus, by mitigating line congestions, present a solution to the fundamental issues behind the control of voltage in power systems~\cite{ORBi-30326279-329a-427d-a9d3-c67606fa5602}.
The potential contribution of the virtual admittance controller (VAC) to the voltage regulation from DG units was briefly explored in~\cite{moutevelis2021virtual}.
In this work, only the effect of one DG was studied.
However, to study effects specific for each DG and its connection point~(e.g., distribution lines mismatch and different loading for each system node) multi-converter system configurations must be taken into consideration.
Also, it is necessary to assess the impact of the controller on the overall system, not just at the DG connection point.
In this paper, a novel, recursive secondary controller that is based on the virtual admittance concept and aims to improve the voltage profile of a distribution network~(i.e., the voltage magnitude in all of its nodes) in which several DGs coexist is presented.
This represents an extension of the work originally proposed in the conference version, in which only the primary VAC was considered with a simple, single-converter configuration. No experimental results were presented~\cite{moutevelis2021virtual}.
The proposed method addresses both under- and overvoltage conditions, two problems that are often considered separately in the literature~\cite{vasquez2009voltage,tonkoski2010coordinated}.
The controller supports the grid voltage with both active and reactive power, taking into consideration the natural coupling between them.
This presents a novel approach compared to the well-established practice of using only reactive power for voltage regulation in inductive grids and active power in resistive ones~\cite{vasquez2009voltage}.
To the extent of the knowledge of the authors, no other secondary controller that affects both active and reactive power in a coupled way for voltage regulation exists in the literature.
By using both active and reactive power to regulate voltage, it makes the solution effective for grids with a wide range of $R/X$ ratios.
The centralised algorithm uses measurements of the operating point of the grid to optimally tune the converters and acts as a supplementary control.
For this reason, high bandwidth and fully reliable communication links are not necessary.
This strategy takes into account restrictions of active and reactive power imposed by the DG hardware and the primary source of energy.
The main contributions are first validated by using detailed simulations performed in Matlab/Simulink.
Then, they are tested in the laboratory test bench based on three 15~kVA voltage source converters (VSCs) acting as DGs, and a 75~kVA grid emulator.
The organization of the rest of the paper is as follows.
In Section~\ref{sec.overview}, an application and control overview is given.
In Section~\ref{sec.control.system}, the adaptation of the virtual admittance concept towards the goal of voltage regulation is explained, along with some practical control considerations.
In Section~\ref{sec.optimproblem}, the main contribution of this work, namely the recursive secondary controller is developed and presented.
Sections~\ref{sec.numresults} and~\ref{sec.expresults} present the simulation and experimental results, respectively, validating the theoretical developments.
Lastly, Section~\ref{sec.conclusion} draws conclusions and outlines future work.

\begin{figure}[!t]
\centering
\includegraphics[width=0.9\columnwidth]{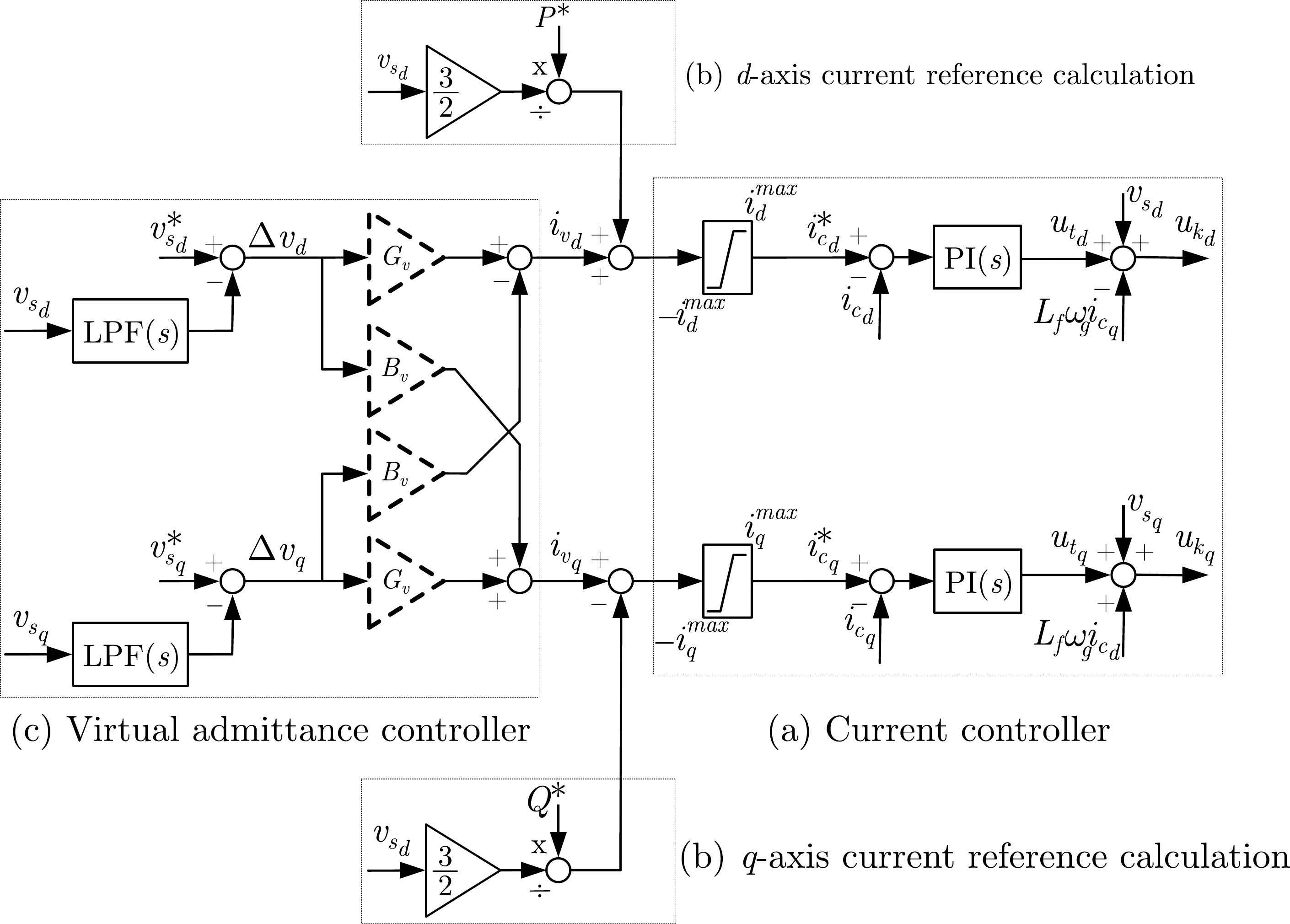}
\caption{Control system of the VSCs. (a) Current controller, (b) current reference calculation and (c) VAC.
Dashed lines indicate that the gain is updated by the secondary controller.}
\label{fig.control}
\end{figure} 
\begin{figure}[!t]
\centering
\includegraphics[width=0.5\columnwidth]{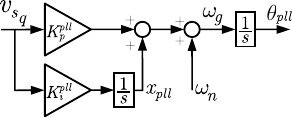}
\caption{Control diagram of the PLL used in the control system of DGs.}
\label{fig.pll}
\end{figure} 
\section{System Overview}
\label{sec.overview}
\subsection{Application Description}
\label{sec.test_system}
Fig.~\ref{fig.grid} shows the electrical diagram of the application studied in this work.
It consists of a radial distribution network with two feeders connected in parallel where the lines have a high $R/X$ ratio.
Radial topology and high $R/X$ ratio for the lines are well known characteristics of low voltage distribution networks.
The latter makes voltage control based purely on reactive power less effective and justifies simultaneous adjustment of active and reactive power for improved voltage regulation.
The selected topology covers the case of a feeder dedicated solely to DG as well as a mixed feeder~(DG and load)~\cite{7457307}.
The other system components are three VSCs, three loads and three interconnection lines.
Each of the converters is connected to the rest of the grid by using a $LCL$ filter (the notation is marked in the figure).
The distribution lines that interconnect node $i$ with node $j$ are modelled by using a $RL$ equivalent, where $R_{ij}$ models the losses and $L_{ij}$ models the inductive nature of the line.
The two feeders are connected to an ideal voltage source, representing the rest of the power network, through another $RL$ circuit ($R_g$ and $L_g$). 
Fig.~\ref{fig.grid} also shows the information flow between the local and secondary control level.
Specifically, voltage and current measurements are collected from the network and power reference set-points as well as synchronization angles are sent by the DGs.
After the secondary level control is executed, VAC gains are sent back to the converters.
The above representative distribution network will be used to demonstrate the practical application of the proposed controller.
Namely, it will be shown how the proposed controller addresses the problem of voltage deviations in active distribution networks by taking advantage of DG controllability.
\subsection{Primary Controller Overview}
\label{sec.control.description}
Fig.~\ref{fig.control} shows the VSC controller that is used in this paper. 
It contains three different parts. 
The first one (Fig.~\ref{fig.control}~(a)) is a standard current controller that is implemented in $dq$ with all required feedback and feed-forward terms, where $\omega_g$ is the frequency estimated by the PLL~\cite{yazdani2010voltage}.
The current controller, PI gains are set based on the pole-placement method considering the dynamics of the converter output $LCL$ filter~\cite{ogata2010modern,yazdani2010voltage}.
Super-script ``*'' stands for ``reference''.
The second one (in Fig.~\ref{fig.control}~(b)) is used to generate the $dq$-axis current references based on the active and reactive power references.
The third one (Fig.~\ref{fig.control}~(c)) is the VAC, which emulates an admittance ($G_v$ and $B_v$) and is used in this work to support the grid voltage.
The low-pass filter in the VAC input guarantees the time scale separation between the two cascaded controllers~(current controller and VAC).
Similar control schemes have already proposed in the literature~\cite{rodriguez2013control,moutevelis2021virtual}.
One should note that the main contribution of this work is not in the controller structure, but in a coordinated method to select the gains of the VAC. 
This is addressed in detail in Section~\ref{sec.optimproblem}.
A control diagram of the PLL that was used in this application for each DG is shown in Fig~\ref{fig.pll}. Its control input is the $q$-axis component of the measured voltage at the connection point of each DG. The PLL in each DG is responsible for its synchronization with the grid.

\subsection{Overview of the Proposed Secondary Controller}
\label{sec.algorithm.description}
\begin{figure}[!t]
\centering
\includegraphics[width=0.99\columnwidth]{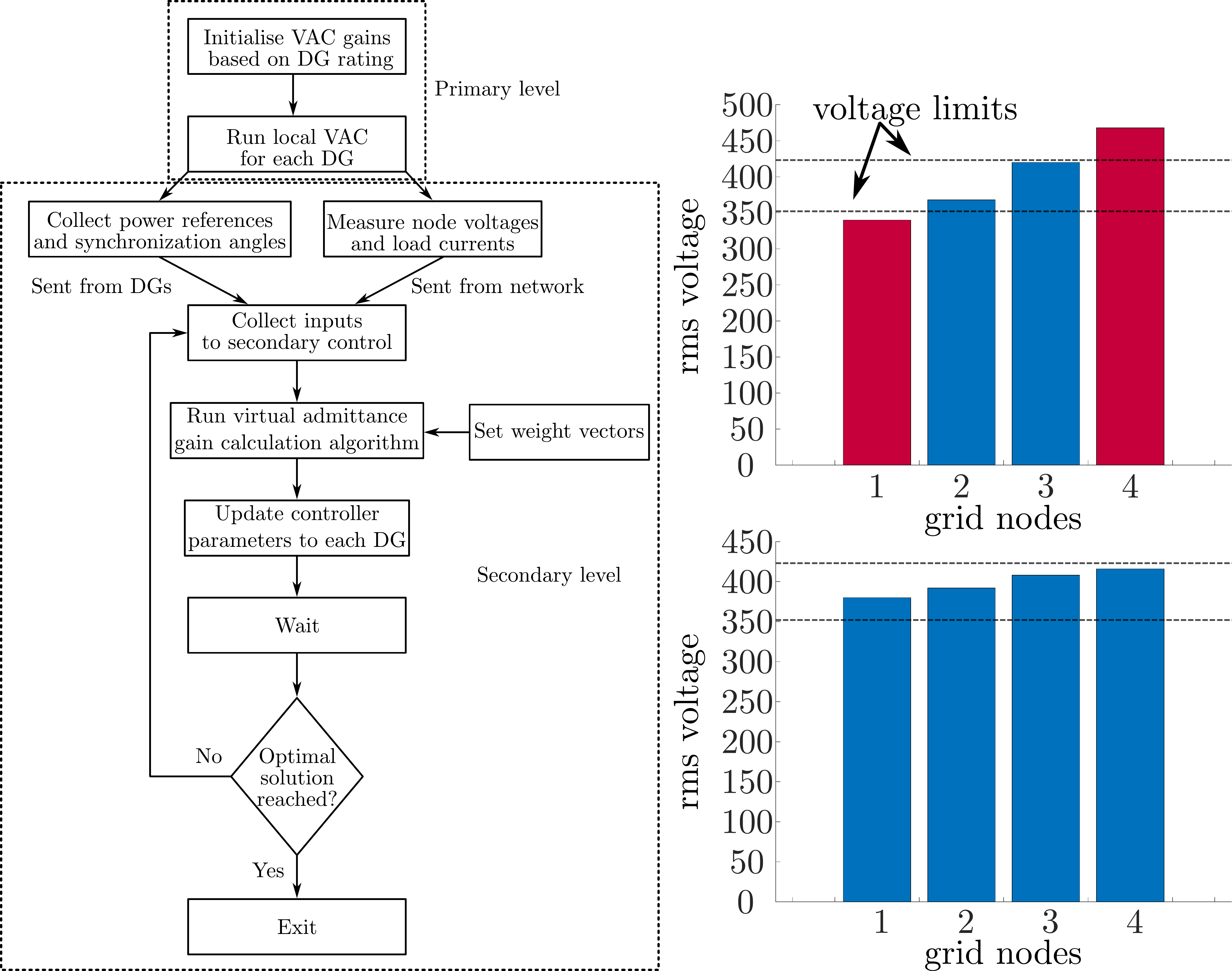}
\caption{(left) Flow diagram with the steps of the proposed controller. 
(right) Example of rms voltages before and after applying the proposed controller.}
\label{fig.flow_ex}
\end{figure}
Fig.~\ref{fig.flow_ex} shows a flow diagram of the algorithm steps and an example showcasing the effect on the distribution grid voltage values. 
First, the primary VAC is enabled to control voltage locally.
The operation of this internal, local controller is described in detail in Section~\ref{sec.control.system}.
Then, the secondary controller is periodically enabled to further regulate the voltages of the distribution grid and to minimize the difference from the nominal value.
The secondary controller is the main focus of this work and its operation is described in detail in Section~\ref{sec.optimproblem}.
In order for the secondary controller to optimally tune the VAC gains in all DGs, measurements for the voltages of each node and current of each load are collected.
The measurements are $dq$ variables that are expressed in a reference frame that is synchronized with the voltage at each node.
For that reason, dedicated synchronization units (PLLs) are assumed.
The synchronization of the DGs is assumed to be achieved by the primary control level and is not further included in the secondary controller as a constraint.
The angles $\delta_i$ are calculated locally and sent from each converter to the secondary controller, as it will be explained in Section~\ref{sec.reference.frames}.
Converters also send to the secondary controller their active and reactive power set-points.
Then, gains are calculated periodically and sent to the converters through a low-bandwidth communication channel.
By measuring network variables and sending them to the secondary control level, any change in the grid operating point is considered by the algorithm in the next iteration.
This allows the secondary controller to optimize the voltage profiles based on the current state of the network.
One should note that the response time of the secondary control level is not critical for the operation of the converters, as the primary level control is supporting the voltage during fast transient events.
For this reason, neither communication delays, computation time nor measurement sampling rate are critical.
In this work, the update period (sampling rate) was selected to be in the order of a few seconds.
The developed algorithm is general and can be applied to different grid topologies.
Although general, the system topology, as well as the line admittances, are assumed to be known.
Although this information may not be readily available for all distribution networks, it is a common requirement for power flow and optimal power flow studies~\cite{evangelopoulos2016optimal}.
\begin{figure}[!t]
\centering
\includegraphics[width=1\columnwidth]{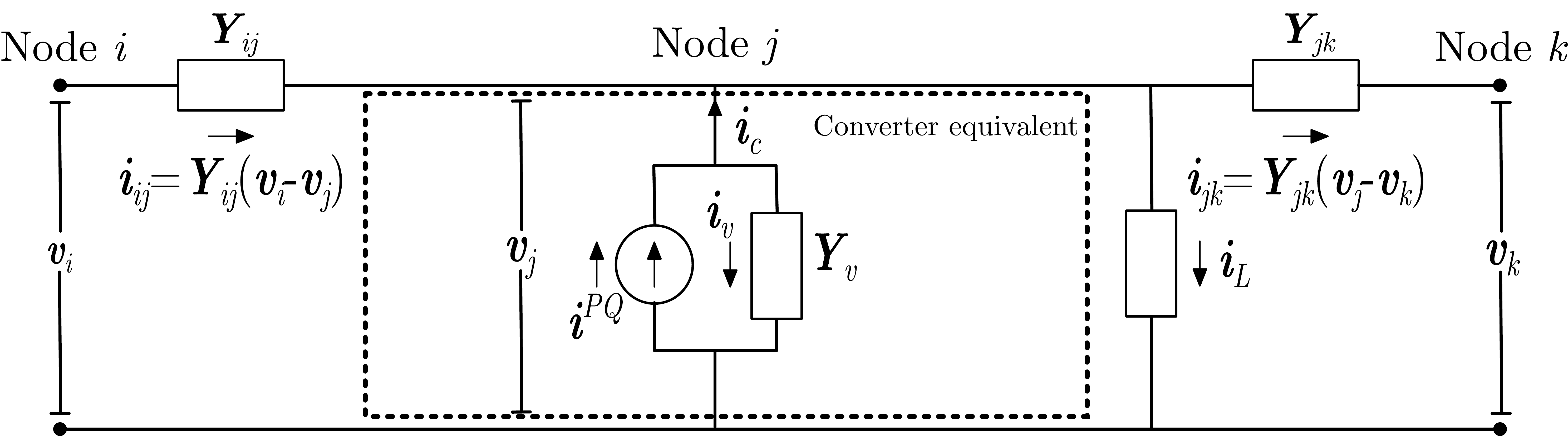}
\vspace{-0.25cm}
\caption{Equivalent steady-state circuit of a VSC with VAC.}
\vspace{-0.3cm}
\label{fig.circuit}
\end{figure}
\begin{table}[t]
\centering
\caption{Comparison between virtual impedance and proposed virtual admittance}
\vspace{-0.1cm}
\footnotesize
\renewcommand{\arraystretch}{1.15}
\begin{tabular}{l|l|l}
\hline
 & Virtual impedance & Virtual admittance \\
\hline
Objective & \makecell[l]{Power sharing \\ Stability} & Voltage profile   \\
Output & Voltage references & Current references \\
Measurements & Current & Voltage\\
Connection & Cascaded connection & Parallel connection\\
DG control type & Grid-forming & Grid-following\\
\hline
\end{tabular}
\vspace{-0.2cm}
\label{tab.virtual_comparison}
\end{table}
\section{Virtual Admittance Control}
\label{sec.control.system}
\subsection{Virtual Admittance Control Modelling}
\label{sec.vir.add.control}
The model of the VAC will be formulated in a $dq$ reference frame that is generated by the PLL in each DG and is rotating with angular speed $\omega_{g}$. The PLL model is:
\begin{equation}  
\label{eq:pll}
\begin{aligned} 
&\Dot{x}_{pll}
=
v_{s_q},\\  
&\Dot{\theta}_{pll}
=
\omega_{g}
=
K_i^{pll} x_{pll} + K_p^{pll} v_{s_q} + \omega_n,
\end{aligned}
\end{equation}
where $K_p^{pll}$, $K_i^{pll}$ are the proportional and integral gains, respectively, of the PLL, $x_{pll}$ is the internal state of the PI controller of the PLL, $\theta_{pll}$ is the angle of the PLL, $\omega_{g}$ is the frequency of the PLL and $\omega_n$ is the nominal frequency of the grid.
Thus, for each DG, an angle $\theta_{pll}$ is generated that is used for the Park transform.

The VAC adds a current component ($i_{{v}_{dq}}$) to the original current reference ($i_{dq}^{PQ}$) generated from the active and reactive power commands ($P^*$ and $Q^*$).
The dynamic equations that comprise the VAC are~\cite{moutevelis2021virtual}: 
\begin{equation}  
\label{eq:virt}
\begin{aligned}
\Dot{i}_{v_{d}}=\frac{v_{s_d}^*-v_{s_d}}{L_{v}}-\frac{R_{v} i_{v_{d}}}{L_{v}}+\omega_n i_{v_{q}},
\\  
\Dot{i}_{v_{q}}=\frac{v_{s_q}^*-v_{s_q}}{L_{v}}-\frac{R_{v} i_{v_{q}}}{L_{v}}-\omega_n i_{v_{d}},
\end{aligned}
\end{equation}
where $v_{s_{dq}}$ are the $dq$ components of the measured voltage at the connection point of the converter, $R_v$ and $L_v$ being the virtual resistance and virtual inductance, respectively. 
The voltage set-point values are $v_{s_d}^*=V_n$ and $v_{s_q}^*=0$, where $V_n$ is the grid nominal voltage.
One should note that the PLL is responsible for setting $v_{s_q}=0$.
Voltage reference value $v_{s_q}^*$ presents an additional control input that can be set arbitrarily.
For consistency between the VAC and the PLL, it was kept at zero.
In the literature, it has been reported that the dynamic VAC can cause synchronous resonances that lead to instabilities~\cite{8693833}.
Therefore, a more robust expression is usually adopted that uses the quasi-stationary approximation~\cite{9570820,9960440}.
By substituting the differential operator ``dot'' in equation~\eqref{eq:virt} with operator $j \omega_n$, one gets:
\begin{equation}  
\label{eq:virt_quasi}
\begin{aligned}
i_{v_{d}}=G_v(v_{s_d}^*-v_{s_d})-B_v(v_{s_q}^*-v_{s_q}),
\\  
i_{v_{q}}=B_v(v_{s_d}^*-v_{s_d})+G_v(v_{s_q}^*-v_{s_q}),
\end{aligned}
\end{equation}
where $G_v$ is the virtual conductance and $B_v$ is the virtual susceptance calculated by the real valued functions:
\begin{equation}
\label{eq:imp2adm}
\begin{aligned}
    G_v
    =
    \frac{R_v}{R_v^2+(\omega_n L_v)^2},\;\;
    B_v
    =
    -\frac{\omega_n L_v}{R_v^2+(\omega_n L_v)^2}
\end{aligned}
.
\end{equation}
One should note that expressions \eqref{eq:virt} and \eqref{eq:virt_quasi} are equivalent in steady state, with only their transient response being different.
The VSC is synchronized with the $d$ component of the grid voltage space vector ($v_{s_q}=0$), and therefore:
\begin{equation}  
\label{eq:pqref}
\begin{aligned}
P^*=\frac{3}{2} v_{s_d} i_d^{PQ}, \;
Q^*=-\frac{3}{2} v_{s_d} i_q^{PQ}.
\end{aligned}
\end{equation}

By calculating the current references $i_{dq}^{PQ}$ that correspond to the power references and adding them to the ones generated by the VAC, the total current references are:
\begin{equation}  
\label{eq:iref}
\begin{aligned}
i_{c_d}^*=
i_{v_{d}}
+
i_d^{PQ}, \;\;
i_{c_q}^*
=
i_{v_{q}}+i_q^{PQ}.
\end{aligned}
\end{equation}

The effect of this controller in steady state can be analysed by using the electrical circuit depicted in Fig.~\ref{fig.circuit}.
For illustration purposes, it is assumed that the current controller loop dynamics and the converter filter dynamics are faster than the outer loop dynamics. 
Assuming a fast internal controller, the contribution of a VSC with a VAC can be summarised as shown in Fig.~\ref{fig.circuit}.
The VSC operates as an admittance in parallel to a current source.
This resembles an operation of a passive element since both active and reactive power injected or absorbed are modified according to the grid voltage level.

The virtual admittance and virtual impedance concepts, as duals of one another, present some similarities but also important differences. 
Both controllers emulate the operation of electrical circuits within the converter control and use the quasi-stationary approximation of the differential operator to improve the robustness of the control.
Also, they both receive voltage references at their input.
Their differences are presented in Table~\ref{tab.virtual_comparison}.
\subsection{Current and Power Restrictions}
\label{sec.current_limit}
Due to the limited capacity of the VSC and the power available from the primary source, a saturation scheme should be used.
In this work, active power is prioritised over reactive power (other alternatives are also possible~\cite{kawabe2017novel}).
Also, the sign of the saturated $q$-axis current should follow that of the original $q$-axis current reference.
The above, based on equation \eqref{eq:pqref}, is formulated as:
\begin{equation} 
\begin{aligned}
i_d^{max}
&=
\min ( I^{max}, P^{max}/u_{dc}),
\label{eq:idqlim} \\
i_q^{max}
&=
\sqrt{(I^{max})^2-(i_{c_d}^{*})^2},
\\
i_q^{min}
&=
- \sqrt{(I^{max})^2-(i_{c_d}^{*})^2},
\end{aligned}
\end{equation}
\noindent where $I^{max}$ is the VSC current limit and $P^{max}$ is the maximum power available from the primary source. 
Active power limiting is important since the primary source may have a limited capacity or it might already be operating at its maximum capacity (e.g. the case of a PV inverter operating at the maximum power point).
Also, converters without storage cannot absorb active power (this limitation does not affect reactive power).
Thus, the upper and lower current limits are:
\begin{equation} 
\begin{aligned}
\label{eq:upper_lower_lims}
0<i_d<i_d^{max}\\
i_q^{min}<i_q<i_q^{max}.
\end{aligned}
\end{equation}

Some remarks:
\begin{enumerate}
\item In low-voltage networks the main concern are overvoltages caused by DGs in weak nodes.
In these cases, reducing the VSC output power is technically possible, although its economical impact should be assessed.
Financial loss from the producer side could be recuperated through compensation mechanisms for active power curtailment~\cite{tomar2020compensation}.
\item For undervoltages in resistive networks active power injection is very effective~\cite{kawabe2017novel}, 
but this would require additional active power from the source.
If the DG is already operating at its maximum capacity ($I^{max}< P^{max}/u_{dc}$), the available power would be fully injected and the margin for reactive power injection would be zero. However, when the power available from the primary source is less than the VSC total capacity ($I^{max}> P^{max}/u_{dc}$), active power is still prioritised but the VSC has available capacity that cannot be covered by the primary source. This frees up converter capacity that can be used for reactive power injection.
\end{enumerate}
\section{Recursive Secondary Voltage Controller}
\label{sec.optimproblem}
In this section, the optimisation algorithm for tuning the VAC of DGs is presented.
First, the system modelling using admittances is presented.
Then, the objective function and the network constraints are explained.
Finally, stability and current limit constraints are considered.
The aim is to finely adjust the controllers based on the collected measurements and the operating state of the network.
Hence, the VAC does not take part in the secondary control directly and its operation is affected indirectly by updates applied to its control parameters.
One should note that the voltage reference inputs of the VACs could also be added as decision variables in the optimization problem.
However, in this work they were excluded due to the increased complexity they would add to the algorithm. 
\subsection{System Modelling by using Rotating Reference Frames}
\label{sec.reference.frames}
All the system equations will be analysed in a common rotating reference frame ($DQ$) that is synchronised with the ideal grid voltage. 
However, converter equations for the i-th converter are expressed in the local reference frame ($dq-i$).
Transformation from the local reference frame to the common one is achieved by using a rotation matrix:
\begin{equation}
\label{eq:transformation}
\begin{bmatrix}
f_D \\
f_Q
\end{bmatrix}
={
\begin{bmatrix}
\cos \delta_i & -\sin \delta_i\\
\sin\delta_i & \;\;\;\cos \delta_i
\end{bmatrix}}
\begin{bmatrix}
f_{d-i} \\
f_{q-i}
\end{bmatrix},
\end{equation}
\noindent where $\delta_i$ is the angle between the reference frames. 
In this work, this angle was estimated based on the current measurements from the converter-side and a Thevenin equivalent of the connection point of each DG~\cite{kundur1994power}.
More details can be found in the literature~\cite{pogaku2007modeling,moutevelis2022bifurcation}.
From here on, all voltages and currents will be expressed in the appropriate reference frame ($DQ$ or $dq-i$).
\subsection{Admittance Modelling}
\label{sec.admittance.modelling}
The optimisation problem is formulated using matrix notation for convenience.
Matrices are expressed in boldface and uppercase, while vectors are expressed in boldface and lowercase (e.g., $\bm{i}=\bm{Y}\bm{u}$).
The virtual admittance of the VSC can be written in matrix form as:
\begin{equation}
\label{eq:adm_matrix}
\bm{Y}_{v}
=
\begin{bmatrix}
G_{v} & -B_{v} \\
B_{v} & \;\;G_{v}
\end{bmatrix}
,
\end{equation}
where $G_v$ and $B_v$ are the outputs of the optimisation problem.
\subsection{Objective Function}
\label{sec.objfun}
Minimizing the difference between the nominal voltage and the actual voltage in the nodes of the distribution network is selected as the main objective.
The reasons for this selection are the following:
\begin{enumerate}
    \item Although the network can still operate with a certain level of voltage deviations from the nominal value, minimizing them is important for improving the power quality of the network~\cite{turitsyn2011options}.
    \item Minimizing voltage deviation also minimizes voltage limit violation~\cite{zhang2020hierarchically}.
    \item Operating the network close to the voltage limits may lead to line disconnections and potential voltage collapse~\cite{simpson2016voltage}.
\end{enumerate}
For this work, minimizing voltage deviation and improving voltage profile will be henceforth used interchangeably.
At the same time, some tuning flexibility for the operator should be provisioned in terms of the importance of each node and of the participation to the voltage regulation for each DG. 
This can be written as follows:
\begin{equation}  
\label{eq:obj_fun_flex}
\begin{aligned}
\min_{
G_{v}, B_{v}} 
\sum_{ j \in C_N}
a_{j}
(V_n-\sqrt{(v_{s_dj})^2+(v_{s_qj})^2})^2
\\
+
\sum_{i \in C_{DG}}
b_i
((i_{v_di})^2+(i_{v_qi})^2)
,
\end{aligned}
\end{equation}
where $v_{s_{dq}j}$ are the $dq$ voltage components of node $j$,
$i_{v_{dq}i}$ are the $dq$ virtual current components of the $i$-th DG connected to node $j$, $C_N$ is the set of all network nodes, $C_{DG}\subseteq C_N$ is the set of nodes with an interconnected DG and $a_{j}$, $b_i$ are weighting factors.

The first term of \eqref{eq:obj_fun_flex} expresses the objective of voltage regulation and the weights $a_i$ can be used to prioritize specific nodes over others.
The second term expresses the total virtual current of each DG and, therefore, the contribution of each DG to the voltage regulation.
When the $dq$ components of the virtual current are non-zero, the deviation between the original current commands $i_{dq}^{PQ}$ and the actual current commands $i_{c_{dq}}^*$ creates a mismatch between the active and reactive power set-points and the delivered power by the DG~(see Fig.~\ref{fig.control} and equation~\eqref{eq:iref}).
Weights $b$ can be selected to increase or decrease this deviation for each DG~(e.g., for economic reasons).
Due to the trade-off in the aforementioned objectives, careful tuning of the weights should be considered.
Namely, clear priority of one objective over the other should be considered.
This way, the secondary controller will not be aiming to fulfill contradicting arguments at the same time.  
To attend to this requirement, a simple tuning rule can be expressed as:
\begin{equation}  
\label{eq:weight_rule}
\begin{aligned}
a_j+b_i
&=
1, 
\\
0<a_j 
&\leq 1,
\\
0 
\leq b_i 
&<
1
,
\end{aligned}
\end{equation}
for the the $i$-th DG connected to node $j$. 
Limit values $0$ and $1$ are excluded for parameters $a_j$ and $b_i$, respectively.
If a DG is required to not participate to the voltage regulation at all, disconnection of the VAC is preferred over using these limit values.
Equation \eqref{eq:weight_rule} also covers the case in which multiple DGs are connected to the same node.
\subsection{Network Constraints}
\label{sec.net.cons}
The network topology and the effect of DG current injection on the voltage are considered through the Kirchhoff current law.
For every node $i$ the Kirchhoff current law can be written as follows:
\begin{equation}  
\label{eq:kirchhoff}
\begin{aligned}
\sum_{i\in K_j}\bm{Y}_{ij}(\bm{v}_i-\bm{v}_j)+\bm{i}_c
=
\bm{i}_{L},\;\;
\bm{Y}_{ij}
=
\begin{bmatrix}
G_{ij} & -B_{ij} \\
B_{ij} & \;\;G_{ij}
\end{bmatrix},
\end{aligned}
\end{equation}
where $K_j$ is the set of all nodes connected to node $j$, vector $\bm{i}_{L}=[ i_{L_d}\; i_{L_q} ]^T$ represents all the lumped loads connected to the node, $\bm{i}_{c}=[ i_{c_d}\; i_{c_q} ]^T$ is the converter current,  $\bm{v}_{j}=[ v_{s_d}\; v_{s_q} ]_j^T$ is the voltage of node $j$ and $\bm{Y}_{ij}$ is the line admittance matrix.
Equation \eqref{eq:kirchhoff} is general and valid for every type of element connected to the node.
If no load or DG are connected to the node, $\bm{i}_L$ or $\bm{i}_c$ can be neglected.
The current flow in this generic node interconnection can be seen in Fig.~\ref{fig.circuit}.
By solving for $\bm{v}_j$, the following result is obtained:
\begin{equation}  
\label{eq:v_f(i,v)}
\begin{aligned}
\bm{v}_j
=
(\sum_{i\epsilon K_j}\bm{Y}_{ij})^{-1}
(\bm{i}_c
-
\bm{i}_L
+
\sum_{i\epsilon K_j}\bm{Y}_{ij} \bm{v}_i
),
\end{aligned}
\end{equation}
where ``$^{-1}$'' stands for inverse matrix.
Note that for generic values of line parameters, matrix $\sum_{i\epsilon K_j}\bm{Y}_{ij}$ should always be invertible~\cite{bazrafshan2017comprehensive}. 

For known loads, \eqref{eq:v_f(i,v)} has two unknowns on the right-hand side, namely $\bm{v}_i$ and $\bm{i}_c$.
In order to obtain both, it would be necessary to solve a power flow problem, resulting in increased complexity.
In order to simplify the process, and to avoid solving a power flow problem as an intermediate step, it was decided to use the measured value of $\bm{v}_i$ (to be called $\hat{\bm{v}}_i$) instead of the one obtained from the power flow.
Using this simplification, the solution of the optimisation problem will not actually be optimal, but an approximation instead.
However, the constraints and the objective function will force the solution to be closer to the optimal value than in the previous step.
After an iterative process, if the algorithm reaches a steady state, then:
\begin{equation}
\hat{\bm{v}}_i = \bm{v}_i.
\label{eq.same.voltages}
\end{equation}
This means that, if the algorithm is able to converge, the simplification will give exactly the same solution as the original problem.
Even though the stability of the iterative algorithm is not proven theoretically, rigorous simulations under various operating conditions and experimental results will show that this is fulfilled in a practical application.
It should also be noted that the variations of $ \bm{v}_i$ are relatively small, and therefore the simplification is well justified.

The total current injected by each converter includes the part related to the power controller plus the part coming from the VAC:
\begin{equation}  
\label{eq:iref_matrix}
\begin{aligned}
\bm{i}_c
=
\bm{i}_v
+
\bm{i}^{PQ},
\end{aligned}
\end{equation}
where each part can be calculated as follows:
\begin{equation}  
\label{eq:iv_ipq_matrix}
\begin{aligned}
 \bm{i}_v
=
\bm{Y}_v
(
\bm{v}_s^*
-
\bm{v}_i
),
 \;\;
\bm{i}^{PQ}
=
\Big[\frac{2}{3}\frac{P^*}{v_{s_d}} -\frac{2}{3}\frac{Q^*}{v_{s_q}}\Big]^T
.
\end{aligned}
\end{equation}
Note that these are the equivalent expressions of \eqref{eq:virt_quasi} and \eqref{eq:pqref}, in matrix form and in steady state.
From substituting equations \eqref{eq:v_f(i,v)}, \eqref{eq:iref_matrix}, \eqref{eq:iv_ipq_matrix} to \eqref{eq:obj_fun_flex} and by replacing unknown voltages $\bm{v}_{i}=[ v_{s_d}\; v_{s_q} ]_i^T$ with measured signals $\hat{\bm{v}}_{i}=[ \hat{v}_{s_d}\; \hat{v}_{s_q} ]_i^T$, one obtains the objective function of the problem with regard to decision variables $G_v$, $B_v$.
\subsection{Stability Constraints}
\label{sec.stab_consts}
To avoid adverse effects on VAC closed-loop stability, $R_v$, $L_v$ should be strictly positive~\cite{gothner2021harmonic}.
This is expressed as:
\begin{equation}  
\label{eq:rv_lv_pos}
\begin{aligned}
R_v
=
\frac{G_v}{G_v^2+B_v^2}
>
0,
 \;\;
L_v
=
\frac{-B_v}{G_v^2+B_v^2} \frac{1}{\omega_n}
>0.
\end{aligned}
\end{equation}
One should note that with constant $\omega_n$ and by using \eqref{eq:imp2adm} and \eqref{eq:rv_lv_pos}, parameter pairs $G_v$, $B_v$ and $R_v$, $L_v$ can always be calculated when one of the two is known.
Thus, they can be used interchangeably.
Constraints \eqref{eq:rv_lv_pos} are trivially satisfied when $G_v>0$ and $B_v<0$.
These are the constant lower and upper bounds for variables $G_v$, $B_v$, respectively.
When instead of strictly positive, variables $R_v$, $L_v$ are required to be greater from some minimum values $R_v^{min}$, $L_v^{min}$, constant lower and upper bounds become quadratic constraints.
This can be expressed as:
\begin{equation}  
\label{eq:rv_lv_quad}
\begin{aligned}
-G_v
+
R_v^{min}
(
G_v^2+B_v^2
)
<
0,
 \\
B_v
+
\omega_n L_v^{min}
(
G_v^2+B_v^2
)
<
0.
\end{aligned}
\end{equation}
%
More information regarding the derivation of these constraints can be found in~\cite{gothner2021harmonic}.

\subsection{Current Limit Constraints}
\label{sec.current_consts}
Current limits of DGs are included as constraints in the optimisation problem.
A simplified version of the constraints discussed in Section.~\ref{sec.current_limit} can be expressed as:
\begin{equation}  
\label{eq:current_consts}
\begin{aligned}
i_{c_d}^2
+
i_{c_q}^2
-
(I^{max})^2
<
0
.
\end{aligned}
\end{equation}
Note that this constraint does not contain priority of $d$-axis current over $q$-axis current, but ensures that the total requested current does not exceed the converter rating.
By combining equations \eqref{eq:iref_matrix} and \eqref{eq:current_consts} it can be shown that its a quadratic constraints with regard to the optimization variables.

\subsection{Limitations of the Proposed Controller}
The limitations of the proposed controller and assumptions made during this work are summarized as follows.
The network is assumed to be operating under normal operating conditions.
For this reason, the virtual admittance at the primary level as well as at the secondary recursive control does not provide any voltage ride through.
For the secondary control level, some low-bandwidth communication platform is assumed.
A suitable metering infrastructure is assumed in order to perform synchronized measurements of node voltages and load currents.
The topology of the distribution network as well as the line admittances are assumed to be known \emph{a priori} by the secondary control level.
On the local DG level, some active power reserve is assumed.
Finally, the stability of the proposed secondary recursive algorithm will not be proven analytically but validated by using rigorous simulations and experimental results.
\begin{table}[!t]
\centering
\caption{Hardware and control parameters of VSCs.}
\begin{tabular}{| l | l || l | l | l |}
\hline
Parameter & Value & Parameter & Value \\
\hline
$S_n$ & 15~kVA & $V_n$ & 400~V\\
\hline
$f_{n}$ & 50~Hz & $f_{sw}$ & 10~kHz \\
\hline
$C_f$ & 8.8~$\mu$F & $v_{dc}$ & 680~V  \\
\hline
$L_{f1}$,$L_{f2}$ & 2.3, 0.93~mH & $R_{f1}$,$R_{f2}$ & 160.6, 64.9~m$\Omega$  \\
\hline
$K_{p}^i,K_{i}^i$ & 5.14, 593.27 & $K_p^{pll} ,K_i^{pll}$ & 0.05, 0.95 \\
\hline
$T_{f1}$ & 100~ms & $T_{f2}$ & 100~ms\\
\hline
\end{tabular}
\label{tab.parameters.conv}
\end{table}
\begin{table}[!t]
\centering
\caption{Distribution grid parameters.}
\begin{tabular}{| l | l || l | l | l |}
\hline
Parameter & Value & Parameter & Value \\
\hline
$S_n$ & 75~kVA  & $V_n$ & 400~V \\
\hline
$L_g$ & 0.25~mH & $R_g$ & 0.08~$\Omega$ \\
\hline
$L_{12}$ & 0.9~mH  & $R_{12}$ & 0.7~$\Omega$ \\
\hline
$L_{23}$ & 1.2~mH  & $R_{23}$ & 1~$\Omega$ \\
\hline
$L_{14}$ & 1.3~mH  & $R_{14}$ & 1.0750~$\Omega$ \\
\hline
\end{tabular}
\label{tab.parameters.distr.grid}
\end{table}
\begin{table}[!t]
\centering
\caption{Loading and generation parameters.}
\begin{tabular}{| l | l || l | l | l |}
\hline
Parameter & Value & Parameter & Value \\
\hline
$P_{L1}$ & 50~kW  & $P_1^*$ & 9~kW \\
\hline
$P_{L2}$ & 10~kW & $P_2^*$ & 12~kW \\
\hline
$P_{L3}$ & 20~kW  & $P_3^*$ & 15~kW \\
\hline
\end{tabular}
\label{tab.loadgen}
\end{table}

\section{Simulation Results}
\label{sec.numresults}
\subsection{Test System Description}
%
The network in Fig.~\ref{fig.grid} was used for testing the proposed control algorithm.
All the VSCs were equipped with the controller of Fig.~\ref{fig.control} and gains $R_v$, $L_v$ were periodically updated based on the process described in Section~\ref{sec.optimproblem}.
Tables~\ref{tab.parameters.conv} and~\ref{tab.parameters.distr.grid} present the control and hardware parameters of the VSCs and the main parameters of the network, respectively.
Grid parameters $R_g$, $L_g$ were selected so that, under nominal conditions, ohmic losses amounted to 3.7~\% of the system rating and the ${R}/{X}$ ratio was 1. 
Line impedance values were selected based on the available impedance ratings of the laboratory. 
Loading and generation parameters are shown in Table~\ref{tab.loadgen}.
They are selected so that both overvoltage and undervoltage appear in different nodes.

PI current controllers have been designed with damping set to 0.7 and settling time set to 7~ms.
PLL bandwidths are set to 60~Hz and their damping to one.
For all cases, the reactive power reference value $Q^*$ is set to zero.
To avoid high frequency interactions and noise, voltage input both at the local VAC level and at the secondary level are filtered with low-pass filters of time constant $T_{f1}$.
Specifically, a filter time constant of 100~ms was selected that approximately amounts to a bandwidth ratio of 12 between the current controller and the VAC.
To avoid large transients during each parameter update by the secondary controller, the gains are passed through low-pass filters with a time constant $T_{f2}=100$~ms.
All simulations were performed with Matlab/Simulink and the SimPowerSystems toolbox.
Realistic effects like semiconductor switching and discrete time controllers were considered.  
\begin{figure}[!t]
\centering
\includegraphics[width=0.99\columnwidth]{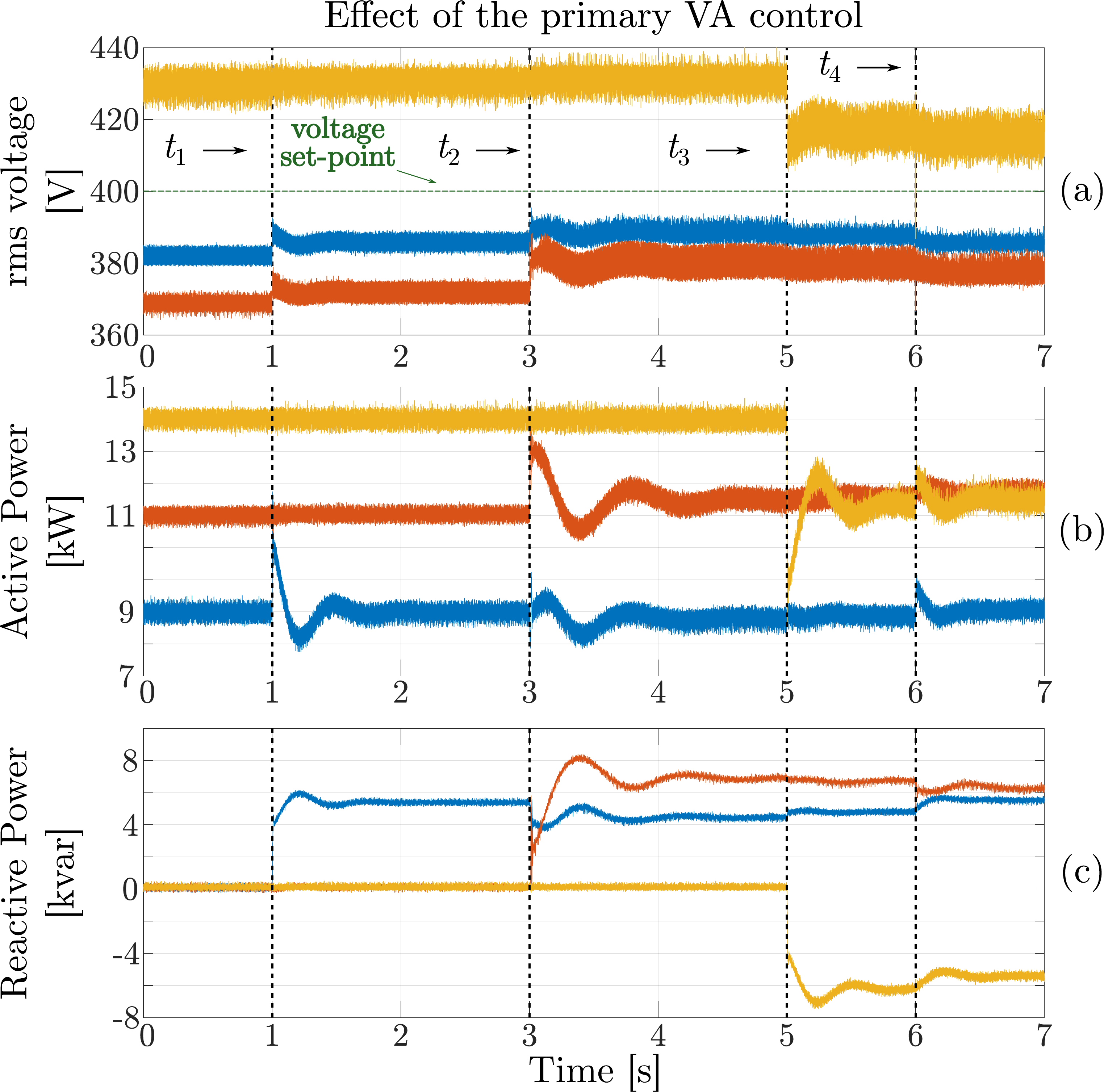}
\caption{Effect of the primary VA control. (a) rms voltage of N2~(blue), N3~(red) and N4~(yellow). (b) Active and (c) reactive power output of DG1~(blue), DG2~(red) and DG3~(yellow) during activation of VAC at $t=1, 3, 5$~s. Load step of 20~kW at Node~1 is performed at $t=6$~s. Voltage set-point is marked with green, dashed line.}
\label{fig.v_PQ1}
\end{figure} 
\begin{figure}[!t]
\centering
\includegraphics[width=0.98\columnwidth]{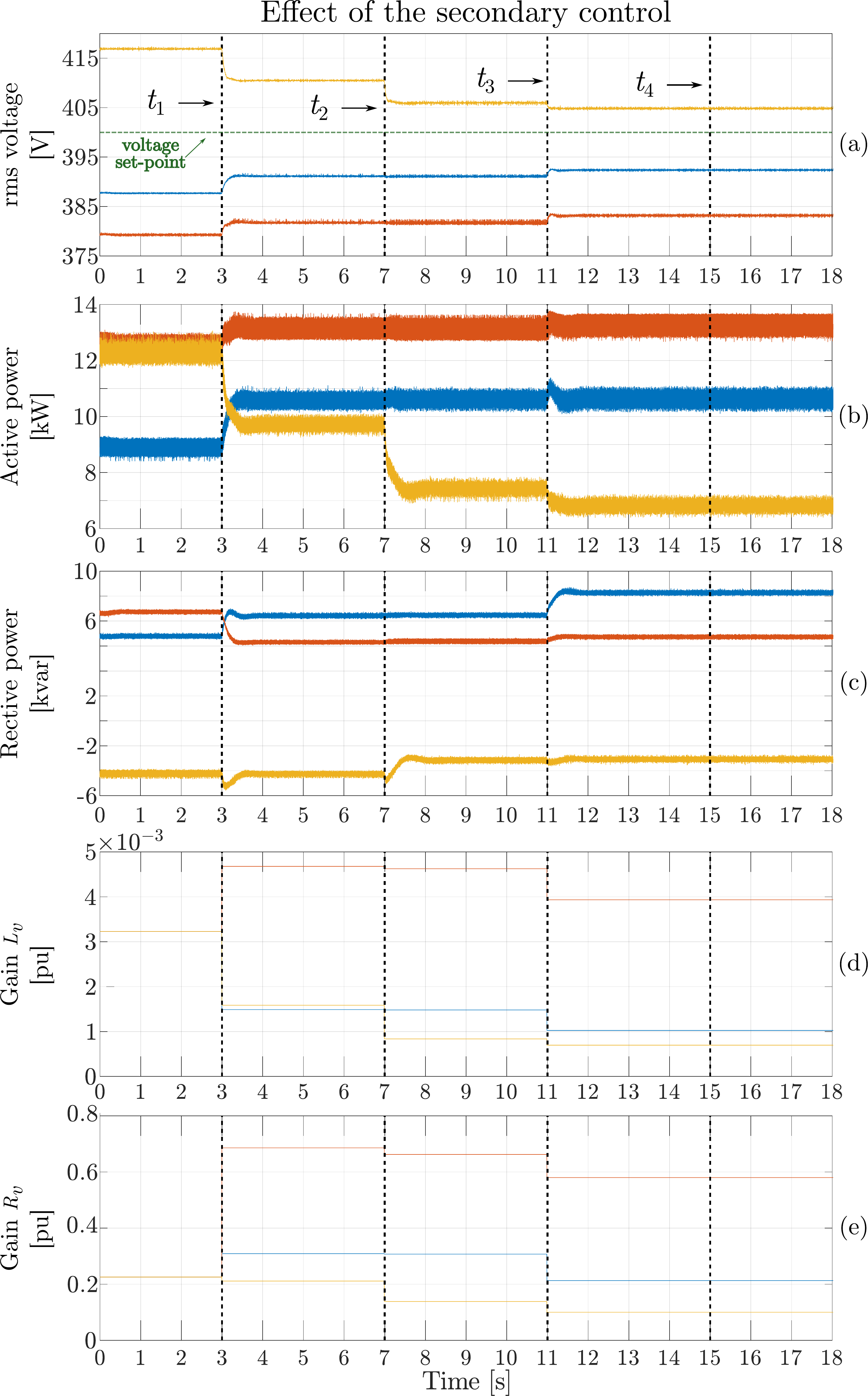}
\caption{Effect of the secondary control. (a) rms voltage of N2~(blue), N3~(red) and N4~(yellow). (b) Active and (c) reactive power output of DG1~(blue), DG2~(red) and DG3~(yellow) during parameter update at $t=3, 7, 11, 15$~s. (d) Gain $L_v$ and (e) $R_v$ of DG1~(blue), DG2~(red) and DG3~(yellow). }
\label{fig.case2}
\end{figure} 
\subsection{Virtual Admittance Primary Control}
\label{prim.control.results}
In this section, the basic functionality of the VAC loop without the parameter update will be demonstrated.
VAC gains are set provisionally to some initial estimates based on the DG ratings
($R_v=0.2255$~pu and $L_v=0.0032$~pu).
Specifically, virtual inductance $L_v$ was set equal to the sum of the rated values of the DG output $LCL$ filter. 
Virtual resistance $R_v$ was set based on the approximate $R/X$ ratio of the $LCL$ circuit. 
For this case, the ratio is estimated between 1.5-4.5~\%.
Fig.~\ref{fig.v_PQ1}(a) shows the impact of the VAC to the rms voltage values of nodes N2, N3 and N4 when the VAC loops are activated at times $t=1,\;3,\;5$~s for DGs 1, 2 and 3, respectively.
Before $t=5$~s there is no direct voltage control for N4.
Its voltage level is kept relatively unchanged by the activation of the VAC loops in N2 and N3 at $t=1$~s and $3$~s, respectively, because N4 is connected to the rest of the network through a dedicated feeder.
Thus, N4 voltage is not benefited indirectly from the voltage control of the other nodes.
Once all the VAC loops are activated, the voltage deviation is reduced both for the case of undervoltage (N2 and N3) as well as for the case of overvoltage (N4).
At time $t=6$~s, a step change of 20~kW is performed at the load connected to N1.
Fig.~\ref{fig.v_PQ1}(b) and ~\ref{fig.v_PQ1}(c) show the changes in the active and reactive power outputs of the DGs.
Initially all the DGs track their respective power references. 
For DG1 and DG2, the activation of the VAC loop leads to an injection of reactive power to support the voltage while the active power output is only slightly modified.
For DG3, reactive power is absorbed and active power is curtailed. 
Both actions are performed automatically to mitigate the overvoltage.
\begin{figure}[!t]
\centering
\includegraphics[width=0.99\columnwidth]{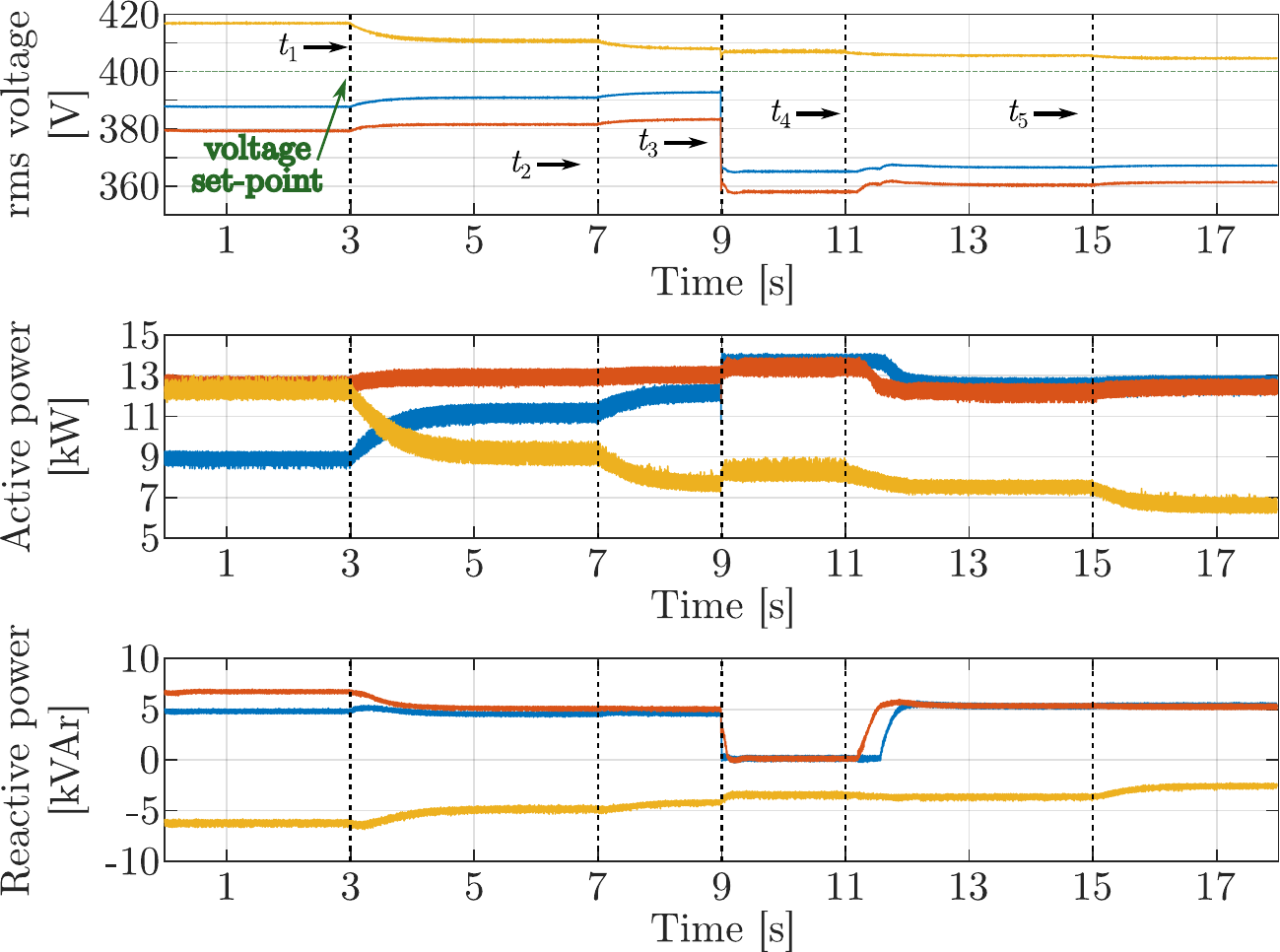}
\caption{Operation of the recursive algorithm during load step change. (a) rms voltage of N2~(blue), N3~(red) and N4~(yellow). (b) Active and (c) reactive power output of DG1~(blue), DG2~(red) and DG3~(yellow) during a 18~kW step increase of the load connected to N2 at $t=9$~s. VAC gain updates at $t=3, 7, 11, 15$~s.}
\label{fig.load_step}
\end{figure} 
\begin{figure}[!t]
\centering
\includegraphics[width=0.99\columnwidth]{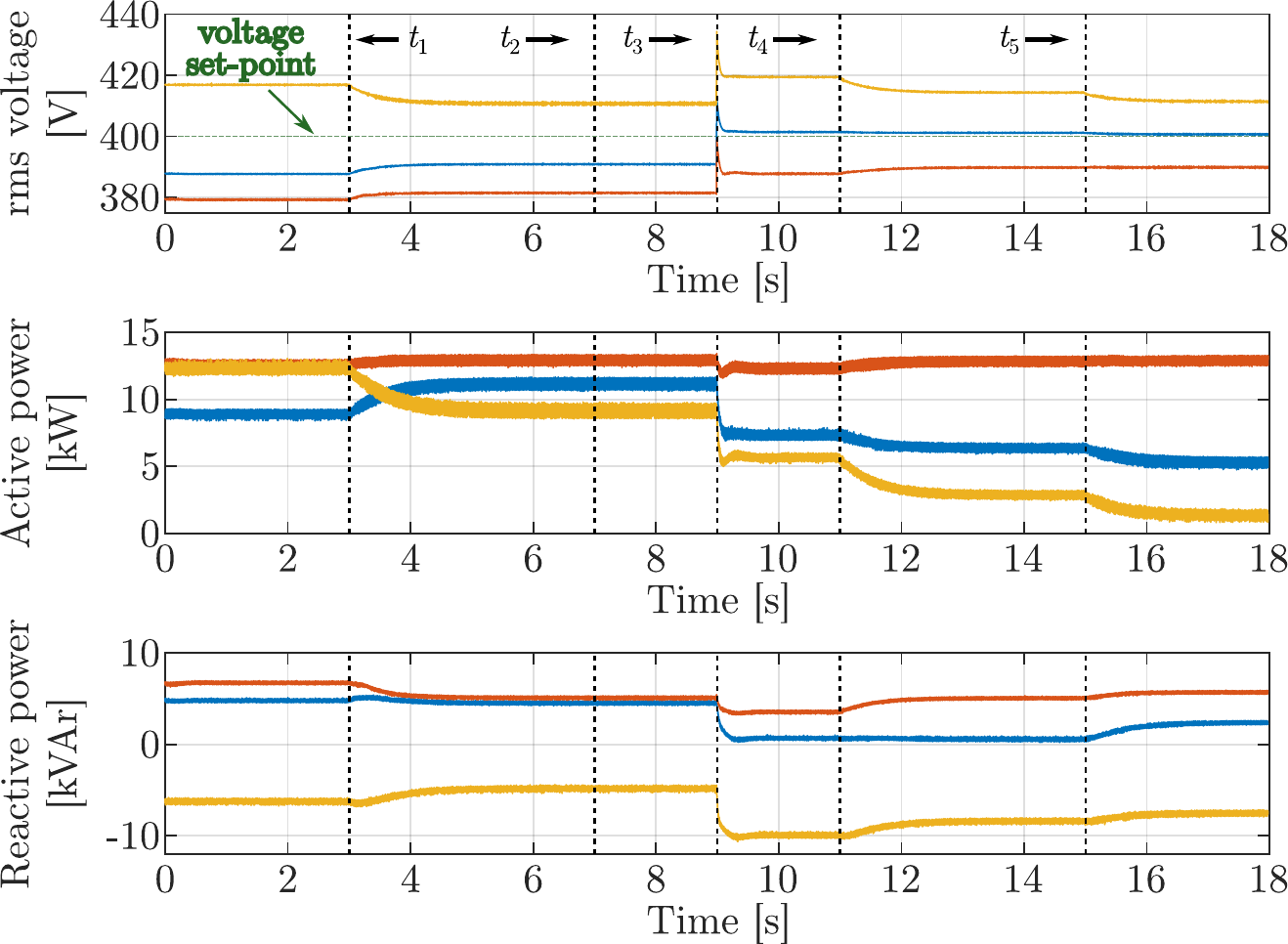}
\caption{Operation of the recursive algorithm during a voltage increase. 
(a) rms voltage of (blue) N2, (red) N3 and (yellow) N4. (b) Active and (c) reactive power output of (blue) DG1, (red) DG2 and (yellow) DG3 during a 7~\% grid voltage step increase of at $t=9$~s. 
VAC gain updates at $t=3, 7, 11, 15$~s.}
\label{fig.voltage_swell}
\end{figure} 
\begin{figure}[!t]
\centering
\includegraphics[width=0.99\columnwidth]{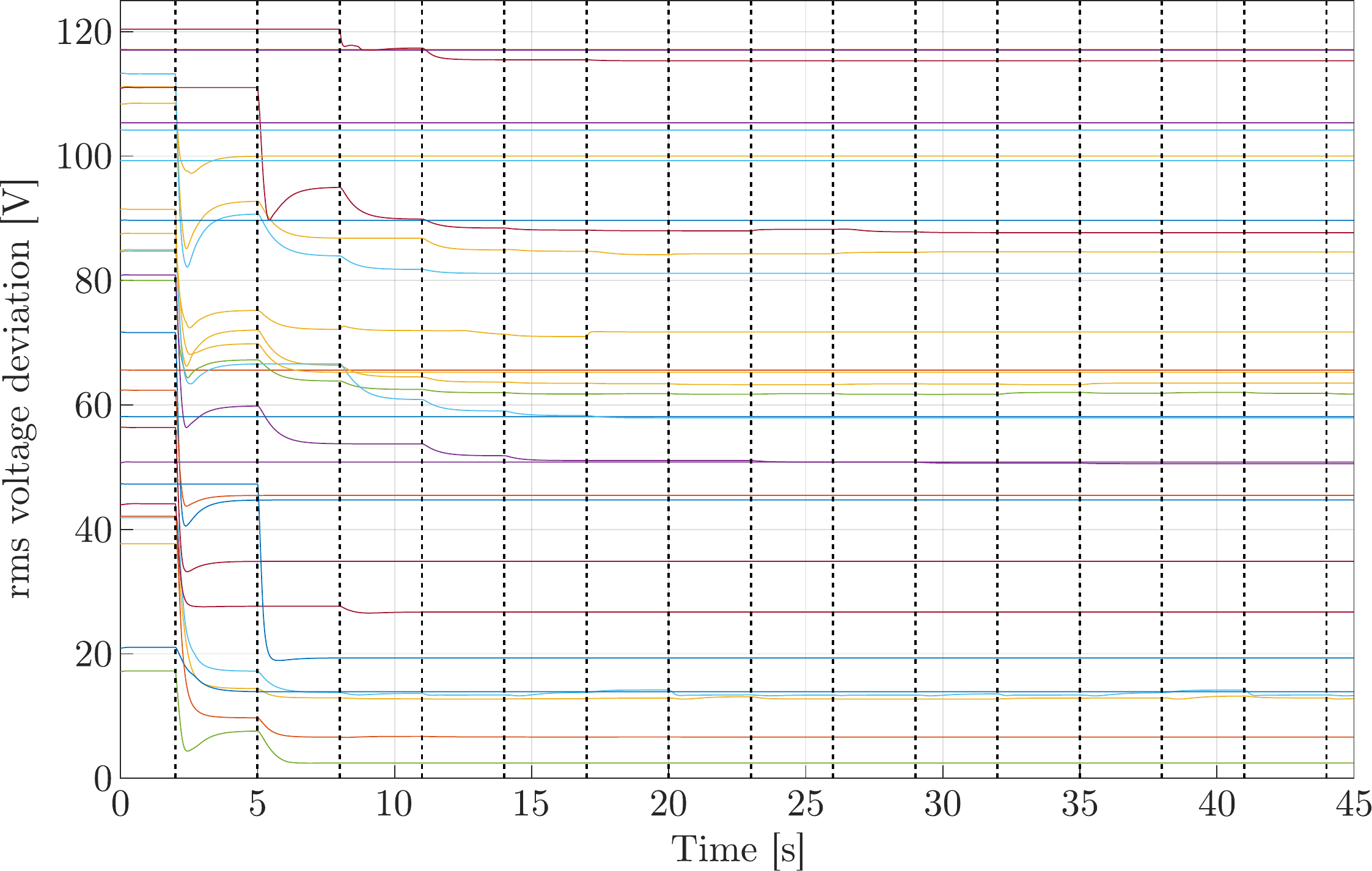}
\caption{rms voltage deviation from the nominal
value for all grid nodes during various loading and generation conditions. Recursive algorithm activated at $t=2$~s and updated every 3~s.}
\label{fig.stoch_objf}
\end{figure} 
\subsection{Secondary Controller}
\label{sec.control.results}
In this section, the operation of the secondary controller is presented.
Initially, all DG units have their VAC activated with gains $R_v$ and $L_v$ set at their initial values of the previous section.
Then, the secondary control algorithm is activated at $t=3$~s and the gains are periodically updated every 4~s.
The minimum values for the gains from equation~\eqref{eq:rv_lv_quad} are set to $R_v^{min}=0.1$~pu, $L_v^{min}=5 \cdot 10^{-4}$~pu.
The weighting factors in the objective function, in \eqref{eq:obj_fun_flex}, are set to $a=[1 \ 1 \ 1 \ 1]$ and $b=[0 \ 0 \ 0]$.
Fig~\ref{fig.case2}(a) shows the effect that the secondary controller has on the voltage of nodes N2, N3 and N4.
It can be seen how both overvoltage and undervoltage are mitigated.
For the purpose of presentation, the voltage values are filtered (the unfiltered values are shown in Fig.~\ref{fig.v_PQ1}).
All the switching effects were fully considered.
Fig.~\ref{fig.case2}(b) and Fig.~\ref{fig.case2}(c) show how the secondary controller affects the active and reactive power output of the DGs.
DG1 and DG2 both increase their active power output to deal with the undervoltage condition.
It should be noted that while the active power requested by the ac part might not be always available from the dc-source, the deviation from the original set-points is small~(10.6~kW instead of 9~kW, and 13.2~kW instead of 12~kW, respectively). 
Hence, exceedingly large power reserves from the dc-side are not necessary.
Reactive power is suitably adjusted as well considering the $R/X$ ratio that the converter ``perceives'', as well as the current limitations.
DG3 curtails active power injection and absorbs reactive power to respond to the overvoltage.

Fig.~\ref{fig.case2}(d) and Fig.~\ref{fig.case2}(e) show the values of the VAC gains during the simulation.
They start from their initial values and are updated at the specified time points.
One should note that the $R_v$ gain of DG3~(yellow line of Fig.~\ref{fig.case2}(e)) reaches its minimum value during the third iteration of the algorithm~($t=11$~s).
This by itself is not a sufficient condition for the recursive algorithm to reach the optimal solution and terminate its operation.
The final iteration of the algorithm~($t=15$~s) outputs the same gains as the previous one, indicating that the recursive assumption \eqref{eq.same.voltages} of the algorithm is fulfilled.
At the same time, the inclusion of constraints \eqref{eq:rv_lv_quad} inside the optimization problem requires that the objective function \eqref{eq:obj_fun_flex} cannot further reduce without violating said constraints.
The simultaneous fulfillment of conditions \eqref{eq.same.voltages} and \eqref{eq:rv_lv_quad}, together with the objective function reaching a value smaller or equal to the previous iteration warrant the optimality of the encountered solution.
Further iterations of the algorithm would output the same gains and the system variables would remain unchanged.
This does not prove the optimality in mathematical terms but strongly suggests that the algorithm is effective in practical applications.

In Fig.~\ref{fig.load_step}, the rms voltage values of nodes N2, N3 and N4 and the active and reactive power outputs of DG1, DG2, DG3 are shown during a 18~kW step increase of the load connected to N2 at $t=9$~s. At $t=3$~s and $t=7$~s the recursive algorithm is executed normally reducing the voltage deviation. At $t=9$~s voltages of N2 and N3 drop instantly due to the load increase.
At $t=11$~s and $t=15$~s the recursive algorithm is executed again regulating the voltage.
In Fig.~\ref{fig.voltage_swell}, the rms voltage values of nodes N2, N3 and N4 and the active and reactive power outputs of DG1, DG2, DG3 are shown during a 7\% grid voltage increase at $t=9$~s. As before, at times $t=3$~s and $t=7$~s the recursive algorithm is executed normally reducing the voltage deviation. At $t=9$~s all node voltages rise due to the grid voltage increase. At $t=11$~s and $t=15$~s the recursive algorithm is executed and the resulting gain update adjusts the active and reactive power outputs accordingly.
The two examples above demonstrate the robustness of the secondary controller against network disturbances and highlight the benefits of the periodic gain updates.

To showcase the robustness of the algorithm under various loading and generation conditions, repeated simulations with stochastically set parameters were performed.
The algorithm was run 30 times with loads getting random values in the range of $[0 \; 50]$~kW and generation getting values in the range of $[0 \; 15]$~kW following a uniform distribution.
Average value models were used for these simulations to speed up computation time.
Fig.~\ref{fig.stoch_objf} shows the total voltage deviation for the whole network and for each iteration. 
In every iteration, the secondary algorithm was executed several times, starting at $t=2$~s and then every 3~s.
For some cases, it should be noted that the objective function does not change throughout the full simulation.
This is because the generation/load mismatch (and hence, voltage deviation) is high and the DGs have reached their saturation limit from the start.
It can be seen that stability was preserved in all cases despite the substantial variation of the network operating points.
\subsection{Effect of Weights}
\label{sec.weights.results}
\begin{figure}[!t]
\centering
\includegraphics[width=1\columnwidth]{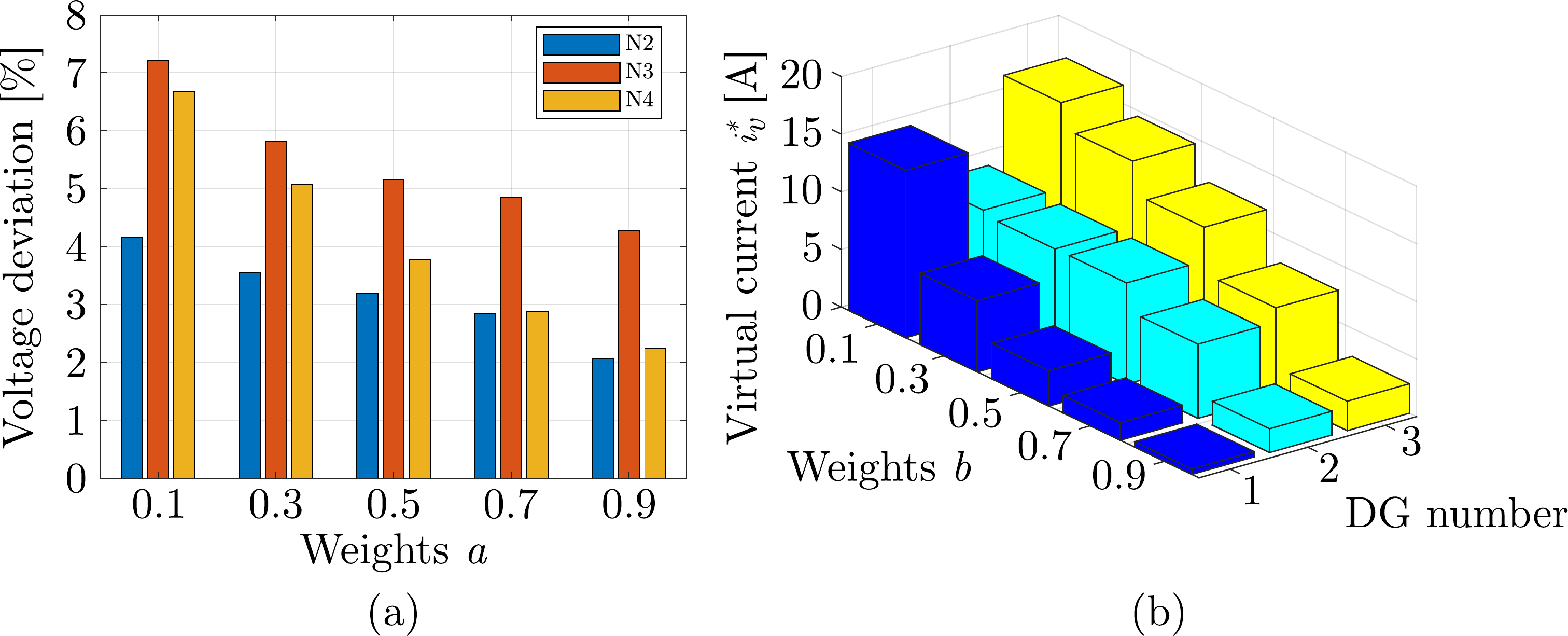}
\caption{Effect of weights - uniform variation. (a) Voltage deviation from nominal voltage at N2, N3 and N4. (b) Virtual current for DG1, DG2 and DG3.}
\label{fig.weights.sym}
\end{figure} 
The effect of the weight vectors $a$ and $b$ in the objective function, in~\eqref{eq:obj_fun_flex}, will be shown. 
The tuning rule presented in~\eqref{eq:weight_rule} was followed.
For the first simulation, weights $a$ were reduced symmetrically for all grid nodes from 0.9 to 0.1 and weights $b$ were increased for all DGs from 0.1 to 0.9.
Steady state values after two parameter updates were used for all the presented variables.
Fig.~\ref{fig.weights.sym}(a) shows the deviation of rms voltages of N2, N3 and N4 from the nominal voltage while Fig.~\ref{fig.weights.sym}(b) shows the total virtual current command $i_v^*=(i_{v_d}^2+i_{v_q}^2)^{1/2}$ for DG1, DG2 and DG3.
The trade-off discussed in Section~\ref{sec.objfun} can be observed. 
For greater values of $a$ (smaller values of $b$), maximum ``effort'' is dedicated from the DGs to the voltage regulation and the deviation decreases with each update.
For small values of $a$ (greater values for $b$), the voltage level improvement caused by the initial effect of the primary VAC is cancelled.
Thus, the deviation from the voltage set-point increases.
The trend of virtual current increasing for larger values of $a$ is not followed by DG2.
This is because DG2 has reached its capacity limit and is not able to contribute to the undervoltage mitigation.
The secondary controller considers the limit and only reduces the virtual current.
\begin{figure}[!t]
\centering
\includegraphics[width=1\columnwidth]{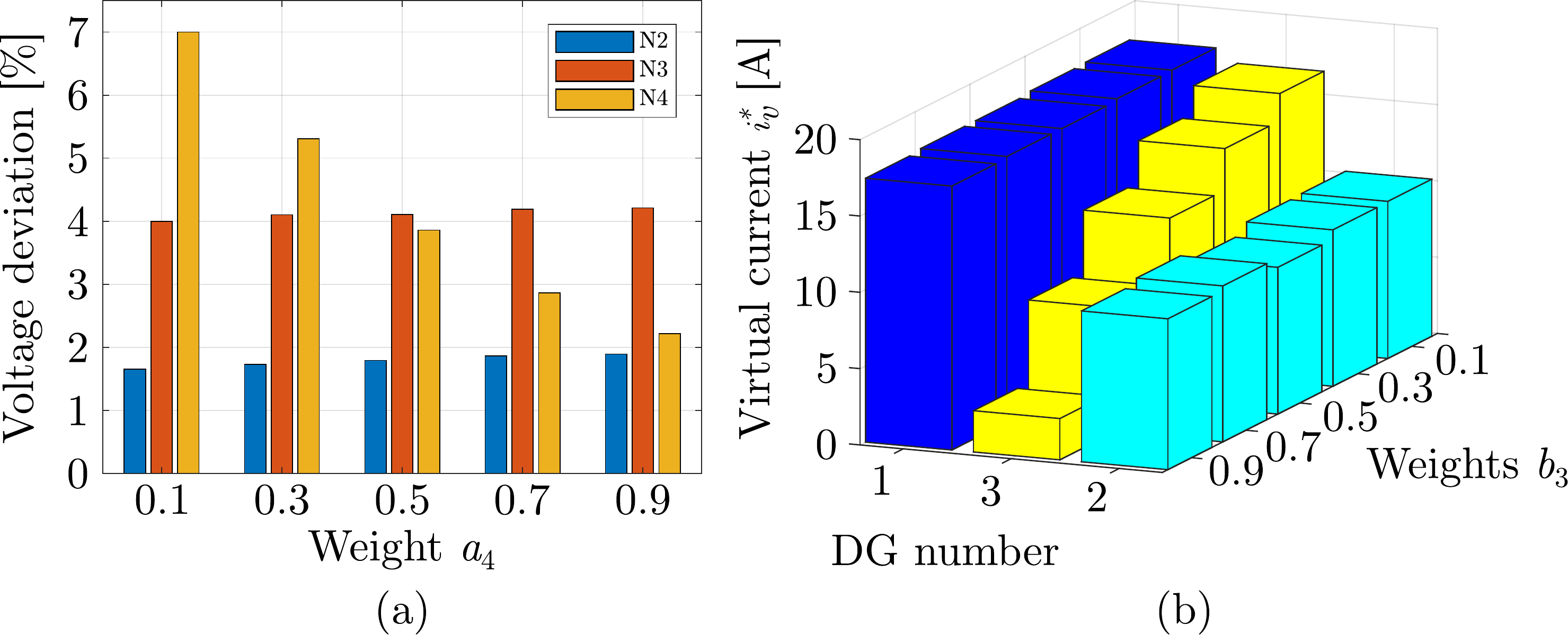}
\caption{%
Effect of weights - variation of $a_4$, $b_3$.
(a) Voltage deviation from the nominal values at N2, N3 and N4. 
(b) Virtual current for DG1, DG2 and DG3.}
\label{fig.weights.asym}
\end{figure} 
\begin{figure}[!t]
\centering
\includegraphics[width=1\columnwidth]{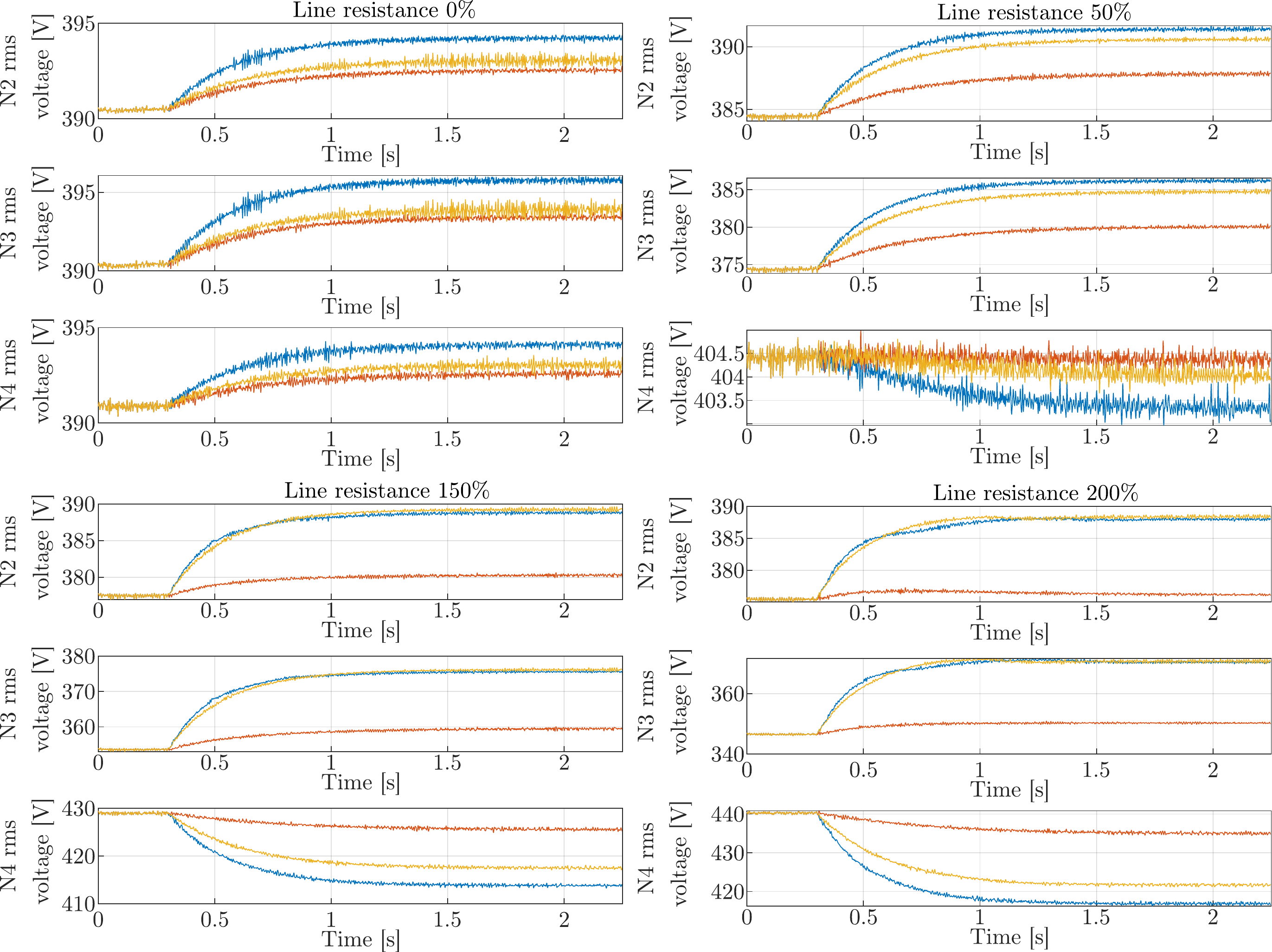}
\caption{rms voltage of N2, N3 and N4. Performance comparison between (blue) VAC, (red) reactive power droop and (yellow) combined active and reactive power droop.
Simulation repeated for different values of line resistances.}
\label{fig.primary_comparison_voltage}
\end{figure} 
Fig.~\ref{fig.weights.asym} shows how the weight selection can individually affect separate nodes or DGs.
Parameter $a_4$ which corresponds to voltage regulation of Node~4 was decreased from 1 to 0.1 (parameter $b_3$ corresponding to converter 3 was increased from 0 to 0.9) while the rest of the weights $a$ were set to 1 (the rest of the weights $b$ were set to 0).
Fig.~\ref{fig.weights.asym}(a) shows that the voltage deviation of N4 follows the same trends as previously, only this time the rest of the nodes remain unaffected.
Fig.~\ref{fig.weights.asym}(b) shows that only the virtual current of DG3 is changed.
For the individual tuning of each weight, consideration of the network topology should be taken.
For example, completely separate regulation of the voltages of Node~2 and 3 is impractical as they are connected to the same feeder. 
\begin{table*}[!t]
\centering
\caption{Summary of control techniques comparison}
\footnotesize
\renewcommand{\arraystretch}{1.15}
\begin{tabular}{l|l|l|l|l|l|l|l}
\hline
Controller & \makecell[l]{Regulated \\ variable} & \makecell[l]{Control \\ level} & \makecell[l]{Communication \\ requirements} & \makecell[l]{Different \\ $R/X$} & \makecell[l]{Over-/ \\ undervoltage} & \makecell[l]{Modified \\ power}  & Performance\\
\hline
$Q/V$ droop & Voltage  & Primary & None & No & Yes & $Q$ & +\\
\hline
\makecell[l]{$P/V$ and \\ $Q/V$ droops} & Voltage  & Primary & None & Yes & Yes & $P$, $Q$ & ++\\
\hline
\makecell[l]{Secondary reactive \\ power controller} & Voltage  & Secondary & \makecell[l]{High bandwidth \\ real-time} & No & No & $Q$ & +\\
\hline
Adaptive droop & \makecell[l]{DG \\ set-point} & Secondary & None & Yes & - & $P$, $Q$ & +\\
\hline
\makecell[l]{ VAC with \\ recursive controller} &   \makecell[l]{Voltage  \\ DG set-point} & \makecell[l]{Primary \\ Secondary} & \makecell[l]{Low bandwidth \\ periodic} & Yes & Yes & $P$, $Q$ & +++\\
\hline
\end{tabular}
\label{tab.comparison_summary}
\end{table*}
\subsection{Comparative Simulations}
\subsubsection{Primary Level} 
To compare the performance of the proposed controller with other methods, simulations were performed at both the primary and the secondary control levels.
For the primary level, the most commonly used technique for voltage regulation is a voltage/reactive power droop~($Q/V$ droop), voltage/active power droop~($P/V$ droop) or a combination of the two~\cite{vasquez2009voltage,tonkoski2010coordinated,collins2015real}. 
Fig.~\ref{fig.primary_comparison_voltage} shows the rms voltages of N2, N3 and N4 when the voltage regulation schemes are enabled at $t=0.3$~s. 
The primary controllers considered were (blue) VAC, (red) $Q/V$ droop and (yellow) combined $P/V$ and $Q/V$ droop.
The simulations were repeated for different values of line resistances. 
For the cases of high line resistances, the performance of the VAC is similar to that of the combined droop when mitigating an undervoltage. 
However, the VAC is more effective for N4, where an overvoltage is observed. 
As expected, the $Q/V$ droop is less effective than the other two techniques that affect both active and reactive power and its efficacy is further reduced for more resistive lines.

\begin{figure}[!t]
\centering
\includegraphics[width=0.85\columnwidth]{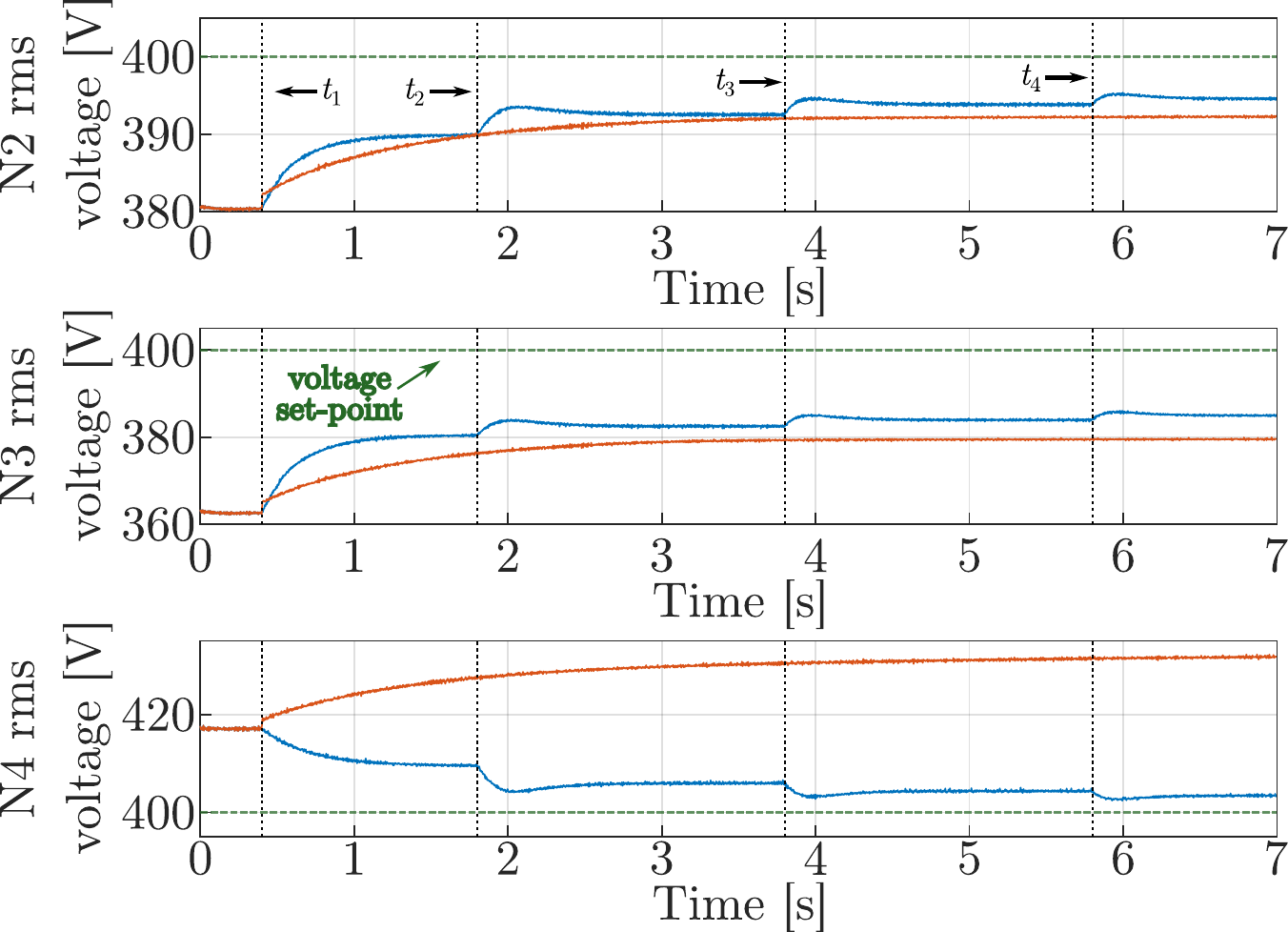}
\caption{rms voltage of N2, N3 and N4. Performance comparison between (blue) the proposed secondary, recursive controller and (red) a centralised, secondary reactive power control. 
At $t=0.4$~s, the primary VAC and the centralised reactive power control are enabled. 
At $t=1.8, 3.8, 5.8$~s the VAC gains are updated by the secondary recursive controller.}
\label{fig.secondary_comparison_node_voltages}
\end{figure} 
\begin{figure}[!t]
\centering
\includegraphics[width=0.85\columnwidth]{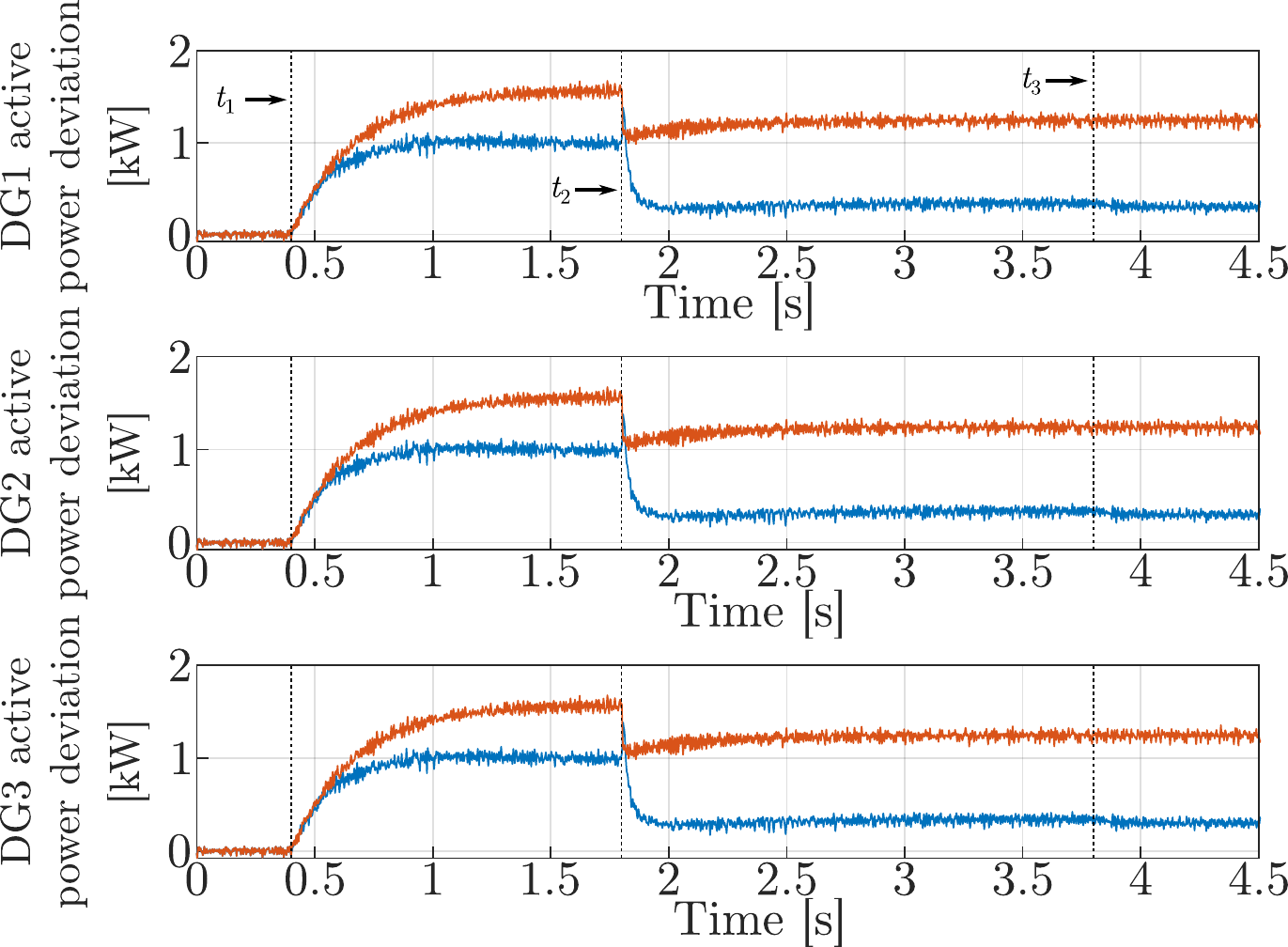}
\caption{Active power deviation for each DG from the active power reference $P^*$. 
Performance comparison between (blue) the proposed secondary, recursive controller and (red) adaptive droop control for active and reactive power. 
At $t=0.4$~s, the primary VAC and active and reactive droop controllers are enabled. 
At $t=1.8, 3.8$~s the VAC gains are updated by the secondary recursive controller. 
At $t=1.8$~s, the active and reactive power droops are updated based on the available active power from the primary source.}
\label{fig.secondary_comparison_dgpower_dev}
\end{figure} 
\subsubsection{Secondary Level} 
At the secondary level, a practical method to control voltage is by adjusting the reactive power set-points of the DGs via a centralised PI controller~\cite{bevrani2017microgrid}.
Then, another method based on an adaptive tuning of the active and reactive power droops was also tested~\cite{li2020adaptive}. 
In the first case, the gains of the secondary reactive power PI controller were set so that the time response of the controller was comparable with the one of the recursive algorithm.
Fig.~\ref{fig.secondary_comparison_node_voltages} shows the rms voltage of N2, N3 and N4 for the two secondary controllers. 
At $t=0.4$~s, the primary VAC and the centralised reactive power control are enabled. 
At $t=1.8,\;3.8,\;5.8$~s the VAC gains are updated by the secondary recursive controller. 
It can be seen that after two VAC gain updates, voltage deviation from the nominal value of 400~V is smaller for the recursive algorithm at N2 and N3.
For N4 and the reactive secondary controller, the reactive power injection causes a rise in the voltage and therefore the profile is deteriorated. 
This is because the secondary algorithm is not designed to address both over- and undervoltages at the same distribution network. 
For the recursive controller, the voltage rise is correctly reduced with each iteration. 
For the adaptive tuning of active and reactive power droops, an update based on the available converter capacity is typically used~\cite{li2020adaptive}.
This droop-tuning method to reduce active power deviation was directly compared with the proposed recursive controller. 
For this comparison, the flexibility of the controller was used and the weights were set to $a=[1 \ 0.1 \ 0.1 \ 0.1]$~($b=[0.9 \ 0.9 \ 0.9]$). 
Fig.~\ref{fig.secondary_comparison_dgpower_dev} shows the active power deviation for each DG from the active power reference $P^*$. 
At $t=0.4$~s, the primary VAC and active and reactive droop controllers are enabled. 
At $t=1.8, 3.8$~s the VAC gains are updated by the secondary recursive controller. 
At $t=1.8$~s, the active and reactive power droops are updated based on the available active power from the primary source.
It can be seen that for the VAC case, active power deviation is smaller for all DGs. 
The performed comparison between the different control techniques is summarized in Table~\ref{tab.comparison_summary}.
\section{Experimental Verification}
\label{sec.expresults}
\subsection{Laboratory Overview}
\begin{figure}[!t]
\centering
\includegraphics[width=0.9\columnwidth]{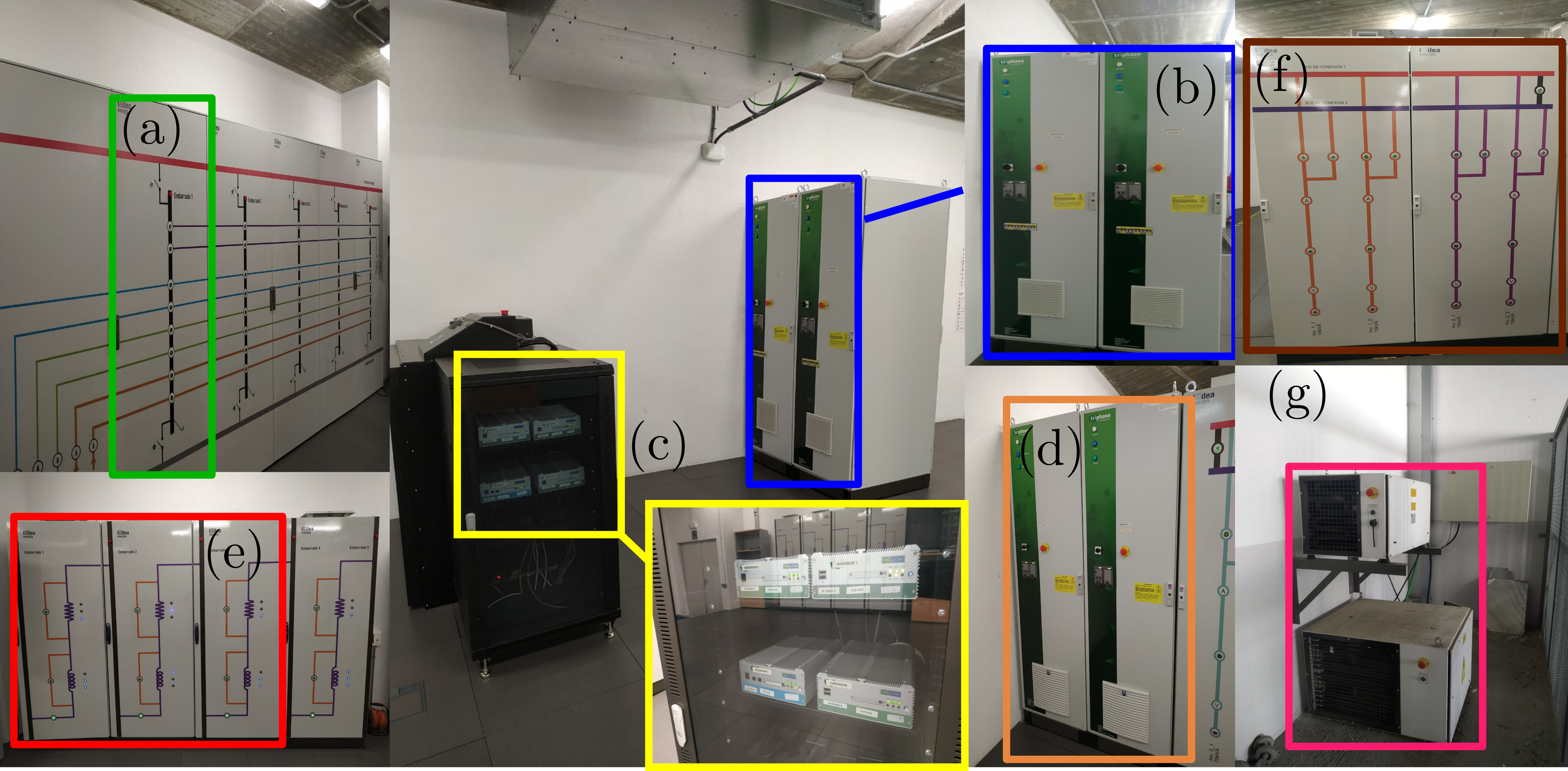}
\caption{Photographs of the laboratory. (a) Ac bus-bars, (b) 15 kVA converters, (c) real-time computers, (d) 75~kVA converters, (e) line impedances, (f) dc bus-bars and (g) passive loads.}
\label{fig.photo_lab}
\end{figure}

\begin{figure}[!t]
\centering
\includegraphics[width=0.97\columnwidth]{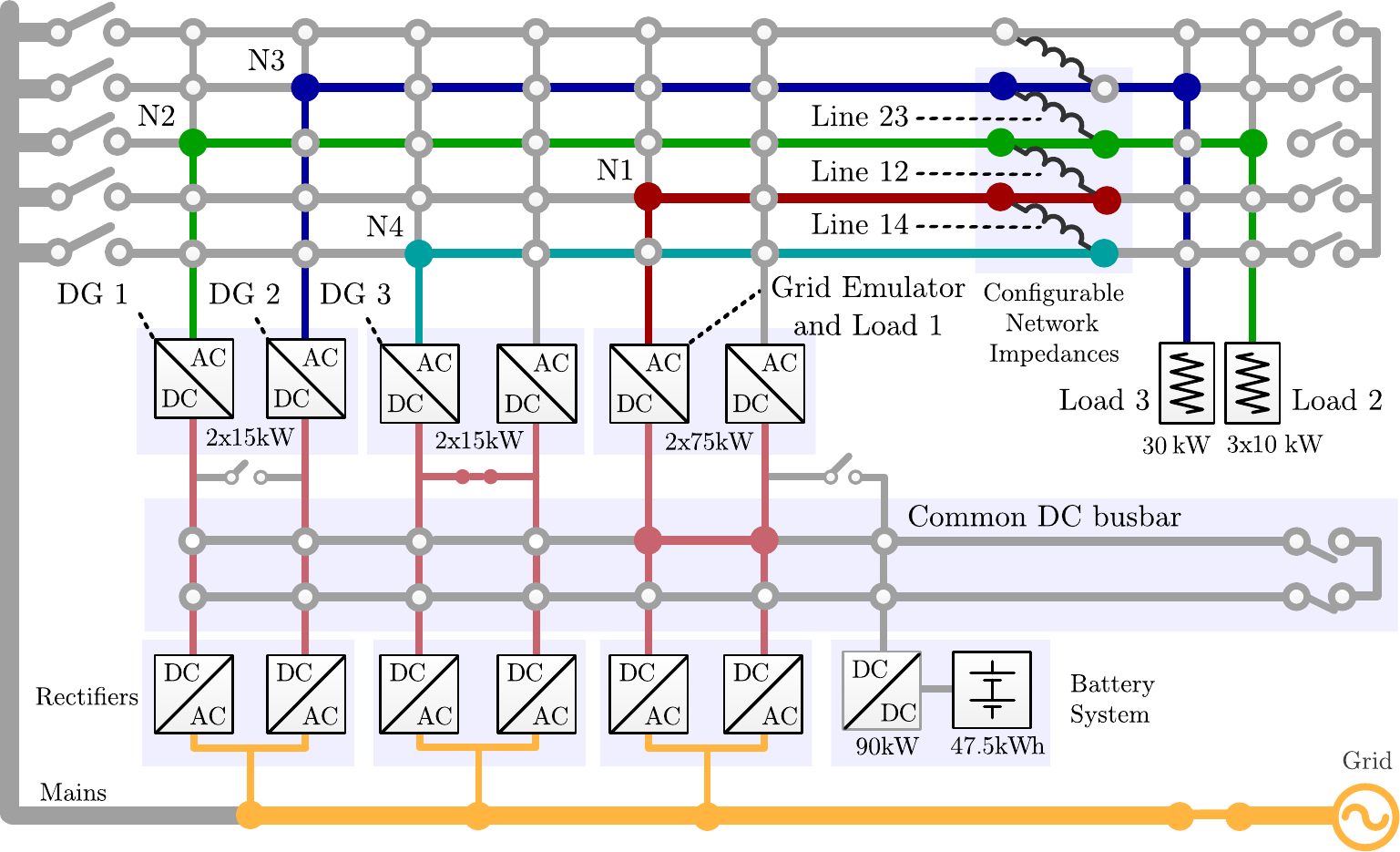}
\caption{Single-line electrical diagram of the laboratory facilities including the test network configuration set-up.
Different colours represent different points of common voltage in the grid.}
\label{fig.lab_schem}
\end{figure}

The functionality of the proposed control structure
was validated in the Smart Energy Integration Lab (SEIL)~\cite{prodanovic2017rapid}.
Fig.~\ref{fig.photo_lab} shows a photograph of the experimental setup used for the experiments while Fig.~\ref{fig.lab_schem} shows the electrical single-line diagram of the lab configuration.
The equivalent lab devices that correspond to the components comprising the test network of Fig.~\ref{fig.grid} are marked. 
Embedded PCs were used for the implementation of the control systems for all converters. 
Two sets of 15~kVA converters were used to represent the DGs of Fig.~\ref{fig.grid}.
The first set was used to represent DG1 and DG2 with the dc voltage being controlled by a set of rectifiers
The second set was used in a back-to-back configuration where one converter was controlling the dc voltage and the other was connected to the test network as DG3.
Converters were connected to the grid by using {\em LCL} filters. 
The sampling and switching frequencies were set to 10 kHz. 
Pulse-width modulation with the min-max modulation strategy and the single update was used~\cite{yazdani2010voltage}.

The grid, the grid impedance and Load 1 were emulated by using a 75~kVA VSC grid emulator. 
The converter that emulated the grid included an $LC$ output filter ($L_f=500$~$\mu$H and $C_f=100$~$\mu$F). 
The filter capacitor was controlled with a fast full state-feedback controller so that the $LC$ filter does not interact with the rest of the system.
The voltage of Node 1 was calculated from the mathematical model (included in the grid emulator) and then used as an input for the voltage controller.
To enable bidirectional power flow from the grid emulator, a different 75~kVA VSC was controlling the dc voltage. 
Their dc-sides were connected through a dc bus-bar.
The distribution lines in Fig.~\ref{fig.grid} were implemented by using a set of configurable impedances with values that are presented in Table~\ref{tab.parameters.distr.grid}.
The entire configuration defines a power hardware in the loop (PHIL) test of the grid defined in Fig.~\ref{fig.grid}. 
\begin{figure}[!t]
\centering
\includegraphics[width=0.92\columnwidth]{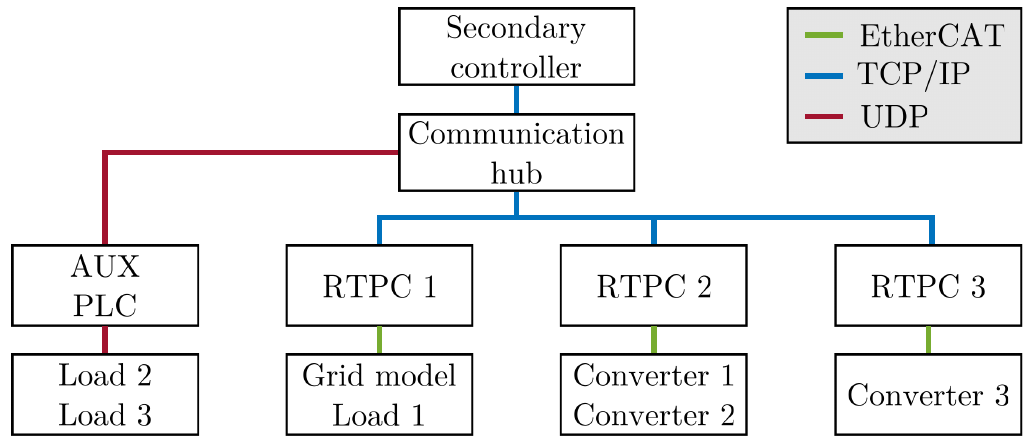}
\caption{Communication infrastructure of the laboratory facilities.}
\label{fig.communication_setup}
\end{figure}
\subsection{Control and Communication Platform Description}
\label{sec.communication}
Fig~\ref{fig.communication_setup} shows a schematic of communication and control setup of laboratory equipment that was used to implement the controllers and data flow depicted in Fig.~\ref{fig.grid}.
The grid and load~1 models were emulated in one real-time PC~(RTPC~1).
The control of converters 1 and 2 was implemented in RTPC~2, while the converter 3 was controlled from RTPC~3.
The control of loads 2 and 3 was implemented in an auxiliary PLC~(AUX~PLC).
The proposed secondary controller was implemented in MATLAB-Simulink and executed in an external PC.
Then, the central communication hub was used to establish a connection and manage a data flow among the devices.
The communication link between the central hub and AUX~PLC was established through UDP protocol, while TCP/IP protocol was used for other devices.
The communication between RTPCs and converters was established via EtherCAT.
This allows execution of internal controllers each sampling period.
\begin{figure}[!t]
\centering
\includegraphics[width=0.97\columnwidth]{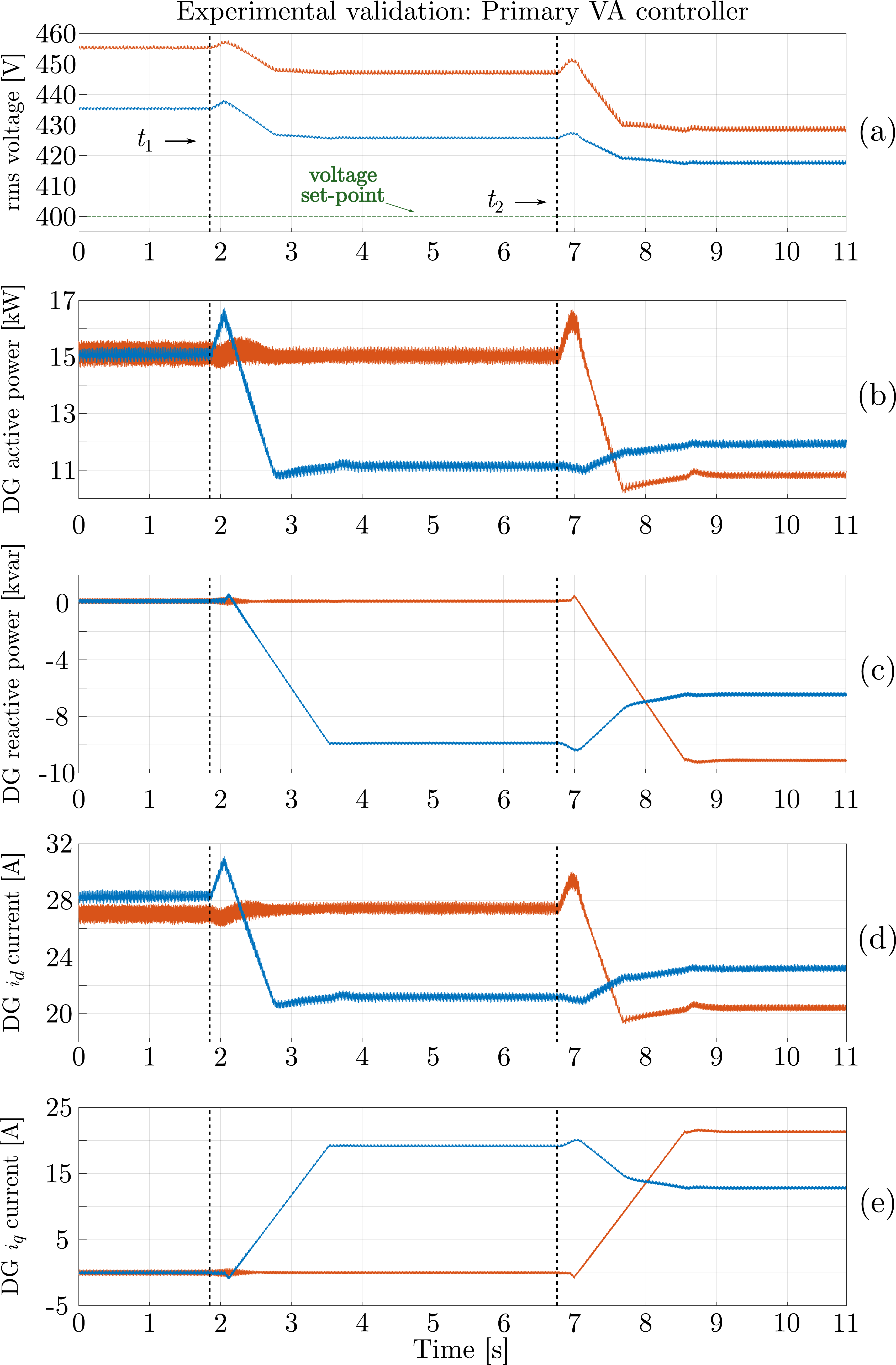}
\caption{Laboratory measurements. (a) rms voltage values for N2~(blue) and N3~(red).
DG measurements for DG1~(blue) and DG2~(red).
(b) Active power.
(c) Reactive power.
(d) $d$-axis current.
(e) $q$-axis current.}
\label{fig.lab_prim}
\end{figure} 
\begin{figure}[!t]
\centering
\includegraphics[width=0.97\columnwidth]{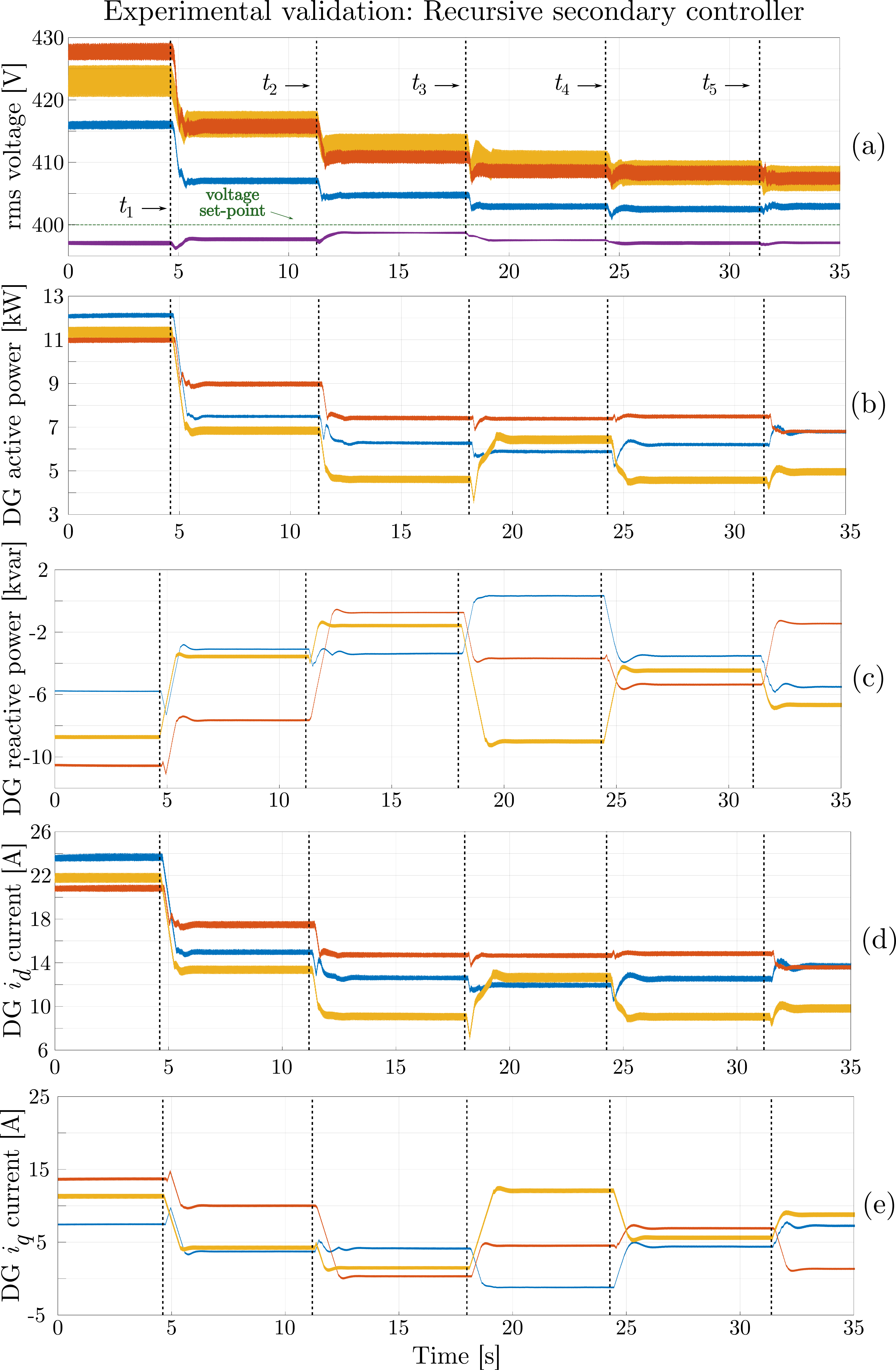}
\caption{Laboratory measurements. (a) rms voltage values for N1~(purple), N2~(blue), N3~(red) and N4~(yellow).
DG measurements for DG1~(blue), DG2~(red) and DG3~(yellow).
(b) Active power.
(c) Reactive power.
(d) $d$-axis current.
(e) $q$-axis current.}
\label{fig.lab_sec}
\end{figure} 
\subsection{Experimental Results}
To validate the operation of the proposed control structure, two experiments were conducted.
First, the basic operation of the primary VAC loop was assessed and its effect on the test network voltage was studied. Then, the recursive secondary controller was activated to tune the VAC gains in real time using the communication platform described in Section~\ref{sec.communication}.
The loading and generation conditions for the laboratory experiments are presented in Table~\ref{tab.loadgen.lab}.
To avoid large transients and for equipment protection, slew rates of current and power commands were saturated to values of 14.14~A/s and 50~kW/s, respectively
\begin{table}[!t]
\centering
\caption{Experimental loading and generation parameters.}
\begin{tabular}{| l | l || l | l | l |}
\hline
Parameter & Value & Parameter & Value \\
\hline
$P_{L1}$ & 10~kW  & $P_1^*$ & 15~kW \\
\hline
$P_{L2}$ & 6~kW & $P_2^*$ & 15~kW \\
\hline
$P_{L3}$ & 6~kW  & $P_3^*$ & 15~kW \\
\hline
\end{tabular}
\label{tab.loadgen.lab}
\end{table}

Fig.~\ref{fig.lab_prim} shows the measurements during the activation of the VAC loops of DG1 and DG2 at times $t_1=1.8$~s and $t_2=6.75$~s, respectively.
Initially, all three DGs were connected and delivering power according to their respective power references.
Then, VAC loops of DG1 and DG2 are activated in close succession.
Fig.~\ref{fig.lab_prim}(a) shows the rms voltage at the nodes to which the DGs are connected, namely N2 and N3.
It can be see how the overvoltage is mitigated and voltage values decrease within acceptable ranges.
The curtailment of active power and absorption of reactive power of the DGs is shown in Fig.~\ref{fig.lab_prim}(b) and Fig.~\ref{fig.lab_prim}(c), respectively.
The change of the DG power output is achieved by the change of the converter current.
The $d$ and $q$-axis current of both DGs are shown in Fig.~\ref{fig.lab_prim}(d) and Fig.~\ref{fig.lab_prim}(e), respectively.
\begin{figure}[!t]
\centering
\includegraphics[width=0.85\columnwidth]{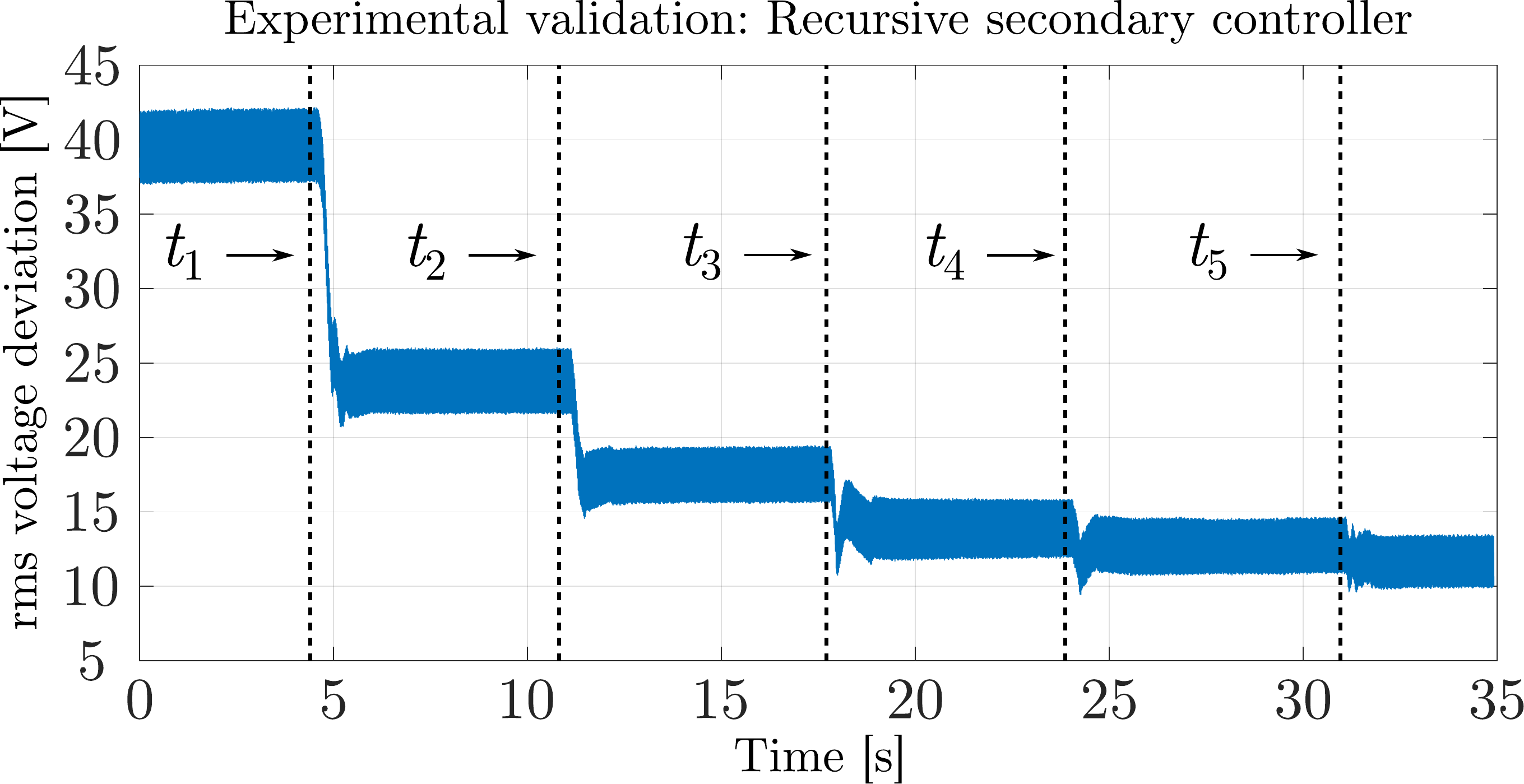}
\caption{Laboratory measurements. rms voltage deviation from the nominal value for all grid nodes.}
\label{fig.lab_sec_obj}
\end{figure} 
\begin{table}[!t]
\centering
\caption{VAC gains during experimental validation of recursive secondary controller.}
\begin{tabular}{| l | l | l | l | l | l | l |}
\hline
\makecell[l]{VAC \\ gain~[pu]}  & \makecell[l]{0~s- \\ $t_1$} & \makecell[l]{$t_1$- \\ $t_2$} & \makecell[l]{$t_2$- \\ $t_3$} & \makecell[l]{$t_3$- \\ $t_4$} & \makecell[l]{$t_4$- \\ $t_5$} & \makecell[l]{$t_5$- \\ 35~s} \\
\hline
\makecell[l]{$L_{v1}$ \\ ($ \cdot 10^{-3})$} & 3.23 & 0.75 & 0.51 & 0.5 & 0.5 & 0.5 \\
\hline
$R_{v1}$ & 0.22 & 0.21 & $10^{-3}$ & $10^{-3}$ & $10^{-3}$ & $10^{-3}$ \\
\hline
\makecell[l]{$L_{v2}$ \\ ($ \cdot 10^{-3})$} & 3.23 & 2.06 & 1.5 & 1.79 & 1.62 & 1.51 \\
\hline
$R_{v2}$& 0.22 & 0.4 & 0.11 & 0.31 & 0.32 & 0.29 \\
\hline
\makecell[l]{$L_{v3}$ \\ ($ \cdot 10^{-3})$} & 3.23 & 1.89 & 1.32 & 1.09 & 0.93 & 0.82 \\
\hline
$R_{v3}$& 0.22 & 0.29 & 0.28 & 0.19 & 0.16 & 0.15 \\
\hline
\end{tabular}
\label{tab.secondary.labtest.vacgains}
\end{table}

For the second experiment, the communication setup described in the previous section was used. The weighting factors were set to $a=[1 \ 1 \ 1 \ 1]$ and $b=[0 \ 0 \ 0]$ while the minimum values for the gains from equation~\eqref{eq:rv_lv_quad} were set to $R_v^{min}=0.001$~pu, $L_v^{min}=5 \cdot 10^{-4}$~pu.
The secondary controller was manually enabled at times $t_1=4.6$~s, $t_2=11.1$~s, $t_3=17.8$~s, $t_4=24$~s and $t_5=31.1$~s.
Fig.~\ref{fig.lab_sec} shows the collected measurements from all grid nodes and all DGs.
It can be seen how the gain updates improve the voltage profiles for both under- and overvoltage.
Fig.~\ref{fig.lab_sec_obj} shows how the total voltage deviation of the network is reduced with each iteration.
One should note that the rate of decrease reduces with each iteration as well, validating the convergence of the algorithm to the optimal solution of the minimization problem.
The VAC gains during the experimental validation of the secondary controller can be seen in Table~\ref{tab.secondary.labtest.vacgains}.
\begin{figure}[!t]
\centering
\includegraphics[width=0.8\columnwidth]{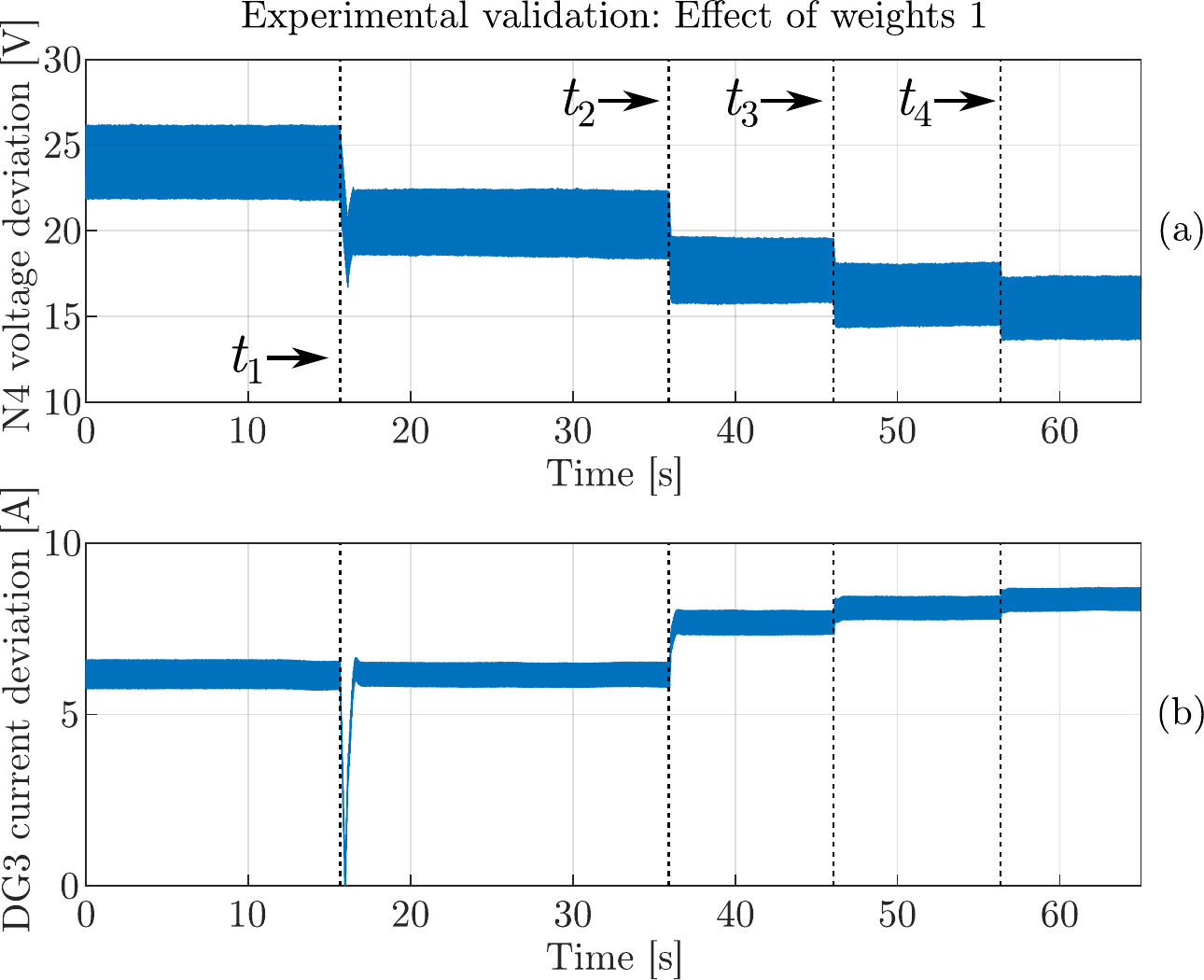}
\caption{Laboratory measurements. (a) Voltage deviation at N4 and (b) Current deviation at DG3.
Objective functions weights set to $a=[1 \ 0.7 \ 0.7 \ 0.7]$ and $b=[0.3 \ 0.3 \ 0.3]$.
}
\label{fig.weights1}
\end{figure} 
\begin{figure}[!ht]
\centering
\includegraphics[width=0.8\columnwidth]{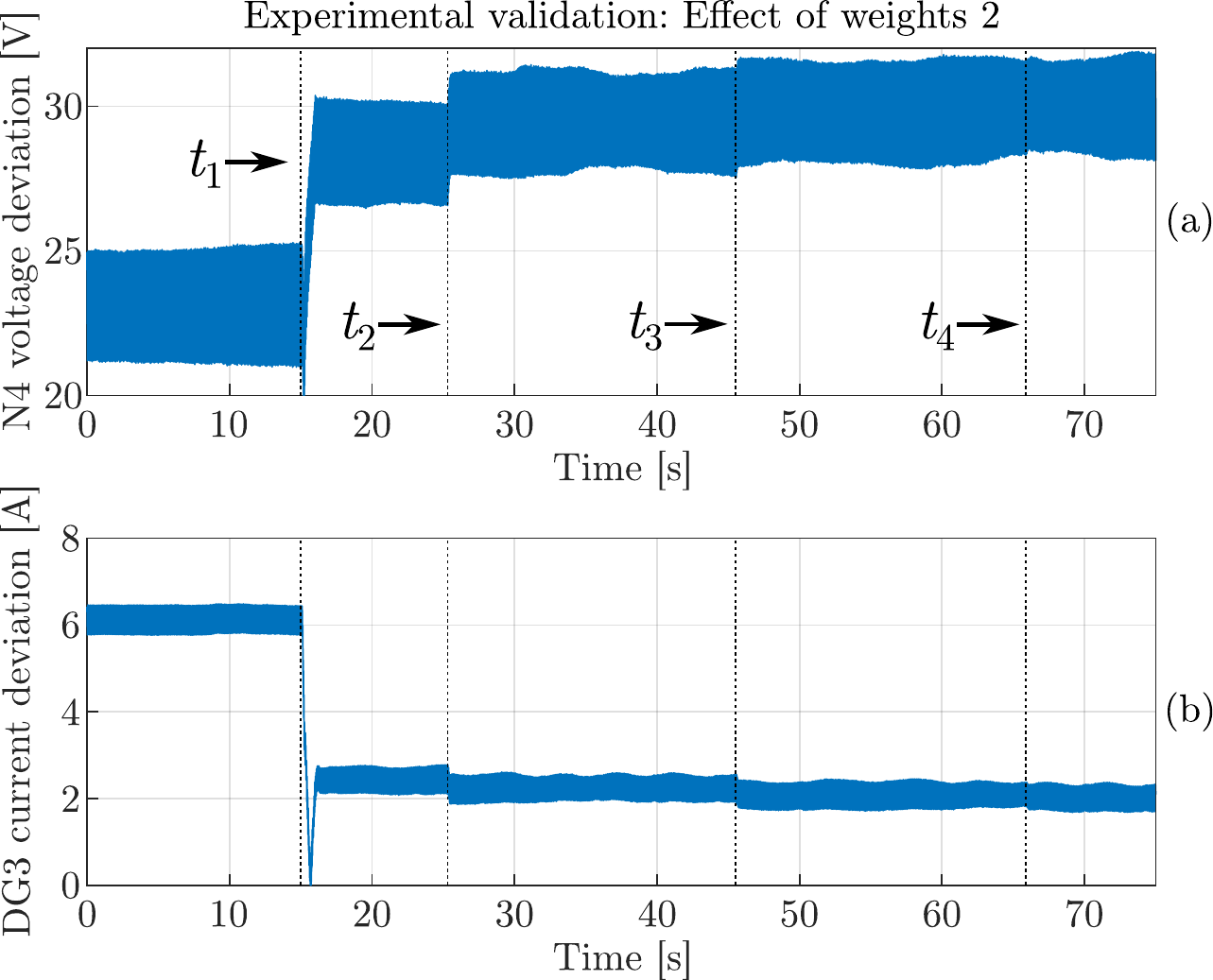}
\caption{Laboratory measurements. (a) Voltage deviation at N4 and (b) Current deviation at DG3.
Objective functions weights set to $a=[1 \ 0.3 \ 0.3 \ 0.3]$ and $b=[0.7 \ 0.7 \ 0.7]$.}
\label{fig.weights2}
\end{figure} 
To validate the functionality of the weighting factors, two experiments were performed.
Fig.~\ref{fig.weights1} shows the voltage deviation at N4 and the total deviation from the nominal current of DG3 when the recursive algorithm was executed with weighting factors set to $a=[1 \ 0.7 \ 0.7 \ 0.7]$ and $b=[0.3 \ 0.3 \ 0.3]$.
The secondary recursive algorithm is executed at times $t_1=15.65$~s, $t_2=35.9$~s, $t_3=46.06$~s and $t_4=56.36$~s.
It can be seen that the voltage deviation is reduced with each iteration. At the same time, the current deviation of the interconnected DG is increased, as the DG puts more ``effort'' towards the goal of voltage regulation.
This is to be expected since priority from the recursive algorithm is given to reducing the voltage deviation of each node.
Fig.~\ref{fig.weights2} shows the same measured variables, this time with weighting factors set to $a=[1 \ 0.3 \ 0.3 \ 0.3]$ and $b=[0.7 \ 0.7 \ 0.7]$.
The secondary recursive algorithm is executed at times $t_1=15$~s, $t_2=25.3$~s, $t_3=45.5$~s and $t_4=65.9$~s.
For this case, reducing the DG deviation from its nominal setpoint is prioritised.
Therefore, the voltage deviation increases while current deviation decreases.
It can be observed that the biggest change occurs during the first iteration of the secondary recursive algorithm.
Following iterations do not affect the operating point significantly as the optimal solution, for this set of weighting factors, has been reached.
The VAC gains for DG3 during the experimental validation of the effect of weights can be seen in Table~\ref{tab.secondary.labtest.weights}.

\begin{table}[!t]
\centering
\caption{VAC gains during experimental validation of the effect of weights.}
\begin{tabular}{| l | l | l | l | l | l |}
\hline
\makecell[l]{Experiment~1 \\ VAC gain~[pu]} & 0~s-$t_1$ & $t_1$-$t_2$ & $t_2$-$t_3$ & $t_3$-$t_4$ & $t_4$-65~s  \\
\hline
$L_{v3}$ ($ \cdot 10^{-3}$) & 3.23 & 1.8 & 1.4 & 1.28 & 1.21  \\
\hline
$R_{v3}$ & 0.22 & 0.87 & 0.7 & 0.6 & 0.55  \\
\hline
\makecell[l]{Experiment~2 \\ VAC gain~[pu]} & 0~s-$t_1$ & $t_1$-$t_2$ & $t_2$-$t_3$ & $t_3$-$t_4$ & $t_4$-75~s  \\
\hline
$L_{v3}$ ($ \cdot 10^{-3}$) & 3.23 & 3.66 & 3.7 & 3.78 & 3.82  \\
\hline
$R_{v3}$ & 0.22 & 1.83 & 2.08 & 2.17 & 2.21 \\
\hline
\end{tabular}
\label{tab.secondary.labtest.weights}
\end{table}
\section{Conclusion}
\label{sec.conclusion}
In this paper, a novel, recursive secondary voltage controller has been proposed based on the virtual admittance concept.
The controller periodically receives measurements from the grid and the DGs and updates the VAC gains by using the proposed recursive optimisation algorithm.
The optimisation objective is to minimize the voltage deviation from the nominal set-point across the grid so that both undervoltage and overvoltage cases are addressed.
Network topology and converter current restrictions are included in the algorithm as constraints.
Weight vectors are also included in the objective function to provide the network operator with the scheduling flexibility. A simple tuning rule for the weight vectors is also provided.
The performance and operation of the secondary voltage controller, as well as the primary VAC, have been assessed and demonstrated by using detailed simulations in a representative low voltage distribution grid.
The flexibility of the secondary controller was compared by using the standard method of adapting the gains of primary control based on the DG available capacity.
Finally, the functionality of the complete control structure was validated through laboratory experiments.
Simulation results have been used to provide the comparison between the proposed controller and other voltage regulation techniques and also to demonstrate its effectiveness.

Simulation results have confirmed that the primary VAC successfully mitigates both under- and overvoltage cases while using exclusively locally measured information.
Then, the secondary controller enables further improvement of the voltage profiles in the grid by updating the VAC parameters.
It has been shown the algorithm converges after a few iterations, thus reaching the optimal solution values for the grid voltage.
The results have also shown that by measuring the network operating point, the secondary controller optimizes the voltage profiles when step changes are applied to different grid variables.
It has been shown the controller maintains stability, even under wide variation of the network load and DG production.
The participation of each DG to the voltage regulation was successfully adjusted by the selection of the weight vectors.
Comparative simulations showed the proposed controller achieves better voltage regulation at the primary level than when only reactive power and combined active/reactive power droops are used (in particular in the overvoltage cases). The controller also shows better results over a range of different $R/X$ ratios.
In the case when DG set-point deviation reduction is prioritised over voltage regulation, the secondary controller achieves better results than the standard method of adapting the values of active/reactive power droops.
Finally, the performed experiments in the SEIL facilities validated the performance of the proposed controller in a realistic application.

Future work will focus on the following topics.
First, theoretical stability analysis of the proposed control will be performed.
Secondly, the parallel operation of multiple VAC-based DGs will be studied.
Finally, the primary source will be modelled in more detail and its effect on the controller performance will be assessed.
\bibliography{arxiv_version}
\bibliographystyle{IEEEtran}

\begin{IEEEbiography}[{\includegraphics[width=1in, height=1.25in, clip, keepaspectratio]{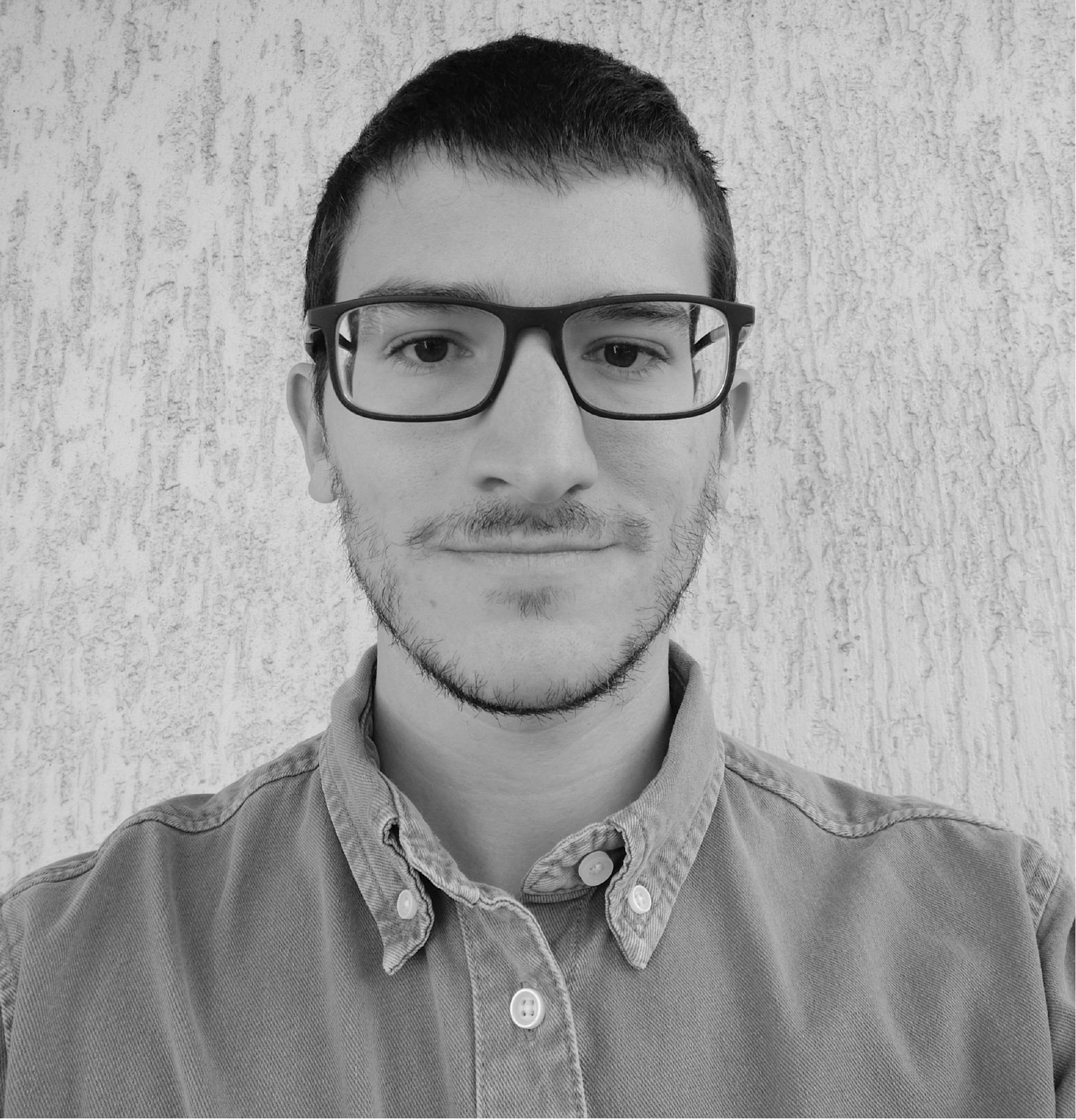}}] {Dionysios Moutevelis} received the M.Eng. degree in Electrical and Computer Engineering from the National Technical University of Athens, Greece in 2017. In 2019 he joined IMDEA Energy Institute, Madrid, Spain where he is currently working as a pre-doctoral researcher. From December 2021, he is also with Alcal\'{a} de Henares University, Department of Electronics, Madrid, Spain. From May to August 2022 he was with University College Dublin, Ireland, as a visiting researcher. His research interests include stability analysis of power systems and power converter control.
\end{IEEEbiography}
\begin{IEEEbiography}[{\includegraphics[width=1in,height=1.25in,clip,keepaspectratio]{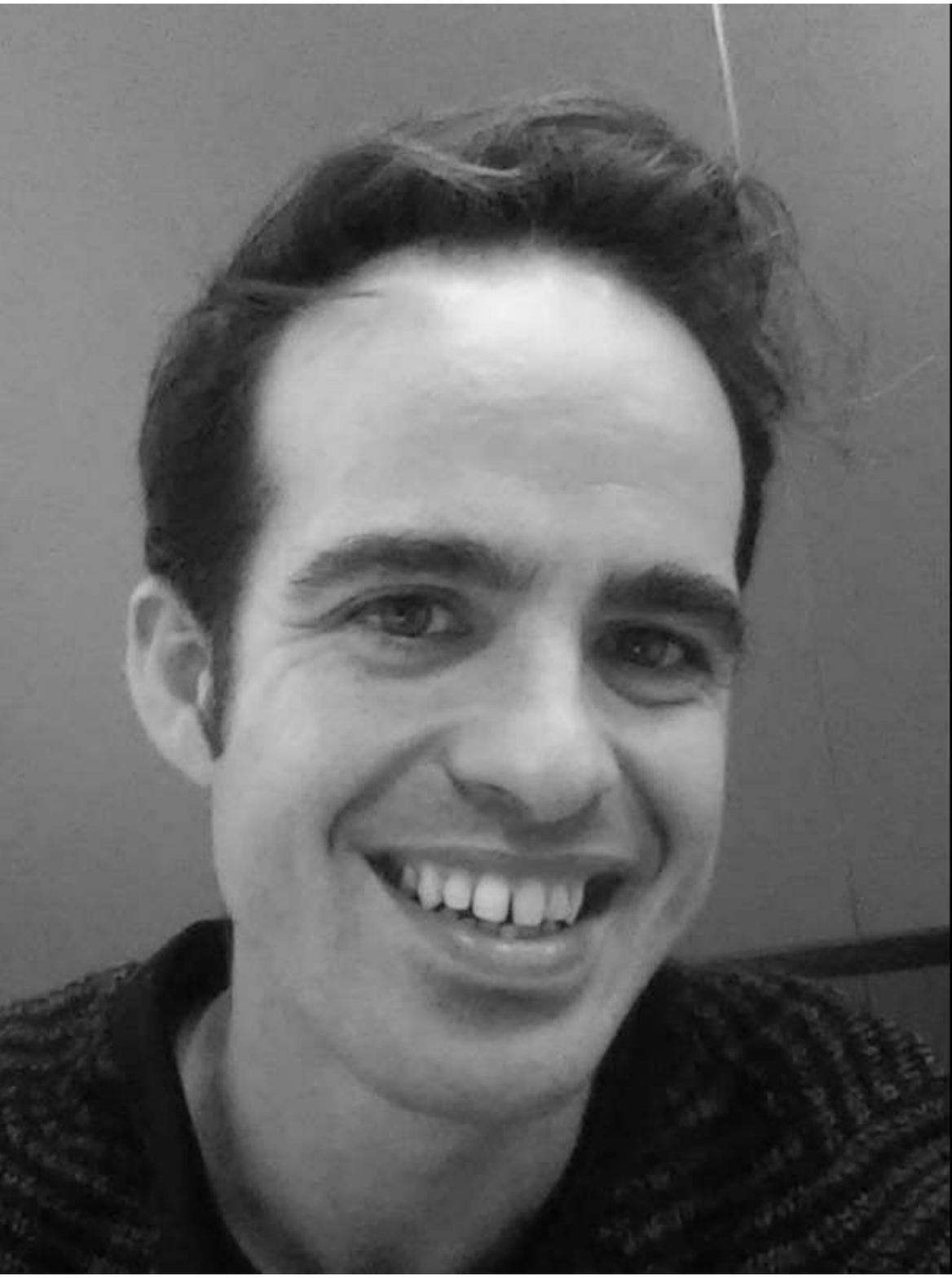}}]%
{Javier Rold\'{a}n-P\'{e}rez} (S'12-M'14) received a B.S. degree in
industrial engineering, a M.S. degree in electronics and control
systems, a M.S. degree in system modeling, and a Ph.D. degree in power
electronics, all from Comillas Pontifical University, Madrid, in 2009,
2010, 2011, and 2015, respectively. From 2010 to 2015, he was with the
Institute for Research in Technology (IIT), Comillas University. In
2014, he was a visiting Ph.D. student at the Department of Energy
Technology, Aalborg University, Denmark. From 2015 to 2016 he was with
the Electric and Control Systems Department at Norvento Energ\'{\i}a
Distribuida. In September 2016 he joined the Electrical Systems Unit at
IMDEA Energy Institute. In 2018, he did a research stay at SINTEF Energy
Research, Trondheim. His research topics are the integration of
renewable energies, microgrids, and power electronics applications.
\end{IEEEbiography}
\begin{IEEEbiography}[{\includegraphics[width=1in,height=1.25in,clip,keepaspectratio]{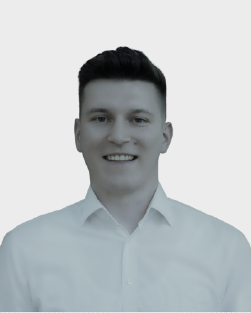}}]%
{Njegos Jankovic} received the B.Sc. and M.Eng. degree in Electrical and Computer Engineering from the Faculty of Technical Science, University of Novi Sad, Serbia in 2016 and 2017, respectively. From 2017 to 2019 he was with Typhoon HIL company, Serbia, developing models of electrical elements. In 2019 he joined IMDEA Energy Institute, Madrid, Spain where he is currently working as a pre-doctoral researcher. From September to December 2022, he was with SINTEF Energy AS Institute, Trondheim, Norway, as a visiting researcher. His research interests include power system stability analysis and the integration of renewable energy sources.
\end{IEEEbiography}
\begin{IEEEbiography}[{\includegraphics[width=1in,height=1.25in,clip,keepaspectratio]{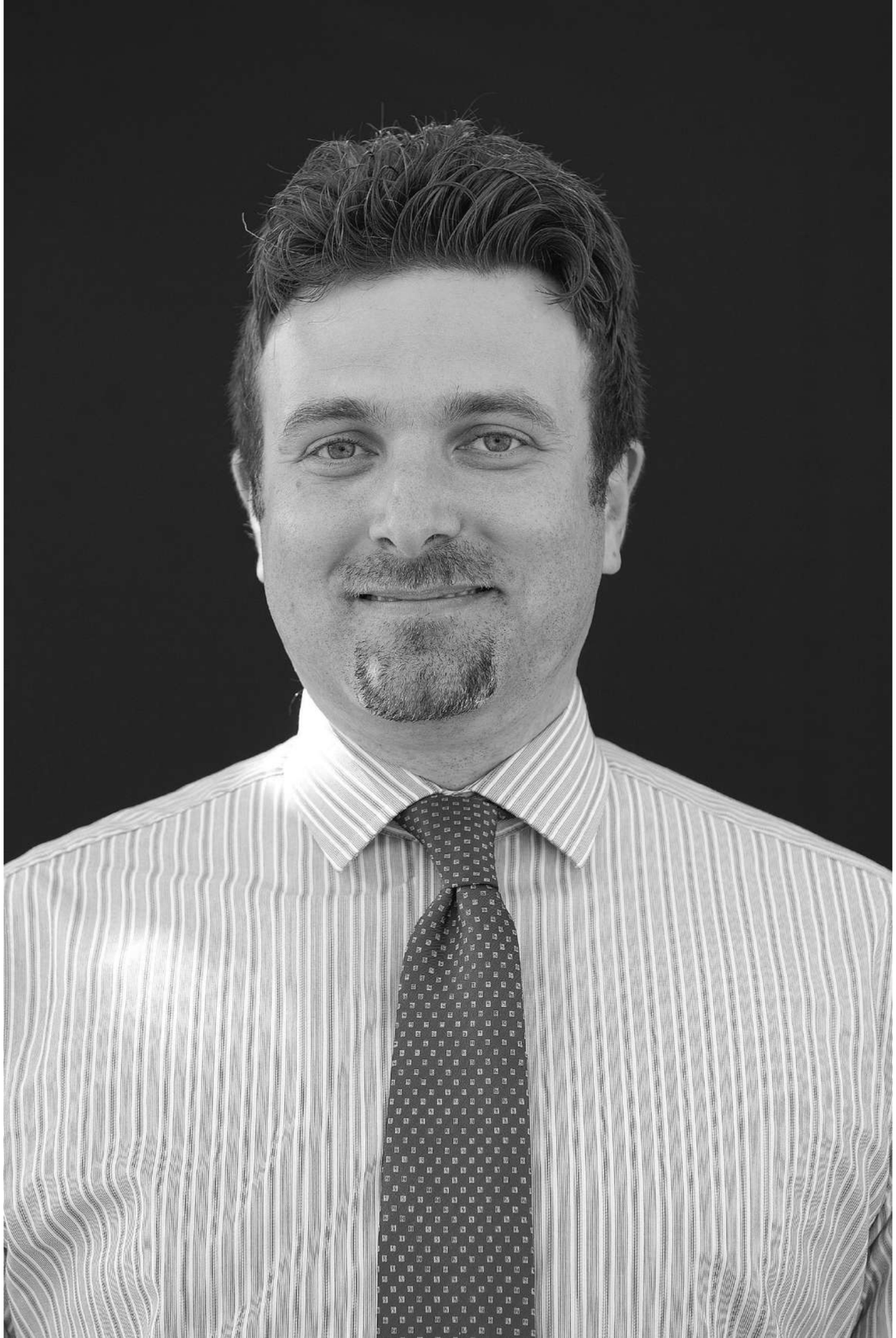}}]%
{Milan Prodanovic} (M'01) received the B.Sc. degree in
electrical engineering from the University of Belgrade, Belgrade,
Serbia, in 1996 and the Ph.D. degree in electric and electronic
engineering from Imperial College, London, U.K., in 2004. From 1997 to
1999, he was with GVS engineering company, Serbia, developing UPS
systems. From 1999 until 2010, he was a Research Associate in electrical
and electronic engineering with Imperial College. He is currently a
Senior Researcher and Head of the Electrical Systems Unit, Institute
IMDEA Energy, Madrid, Spain. He authored a number of highly cited
articles and is the holder of three patents. His research interests
include design and control of power electronics interfaces for
distributed generation, microgrids stability and control, and active
management of distribution networks.
\end{IEEEbiography}
\vfill

\end{document}